\documentclass[%
 reprint,
superscriptaddress,
nofootinbib,
 amsmath,amssymb,
 aps,
]{revtex4-2}

\usepackage{graphicx}% Include figure files
\usepackage{dcolumn}% Align table columns on decimal point
\usepackage{bm}% bold math
\usepackage{epsfig}
\usepackage{epsf}
\usepackage{mathtools}
\usepackage{indentfirst}
\usepackage{amsmath}
\usepackage{CJK}
\usepackage{amsfonts}
\usepackage{subfig}
\usepackage{rotating}
\usepackage{graphicx}
\usepackage{multirow}   
\usepackage{ragged2e}
\usepackage{graphicx}
\usepackage{threeparttable} % For footnotes in tables
\usepackage{amssymb}
\usepackage{booktabs} % in preamble
\usepackage{multirow} % in preamble
\usepackage{float}
\usepackage{placeins}

\usepackage{comment}
\usepackage[separate-uncertainty,retain-explicit-plus, per-mode = symbol, detect-all=true, range-units=brackets, tight-spacing=true]{siunitx}
\usepackage[dvipsnames]{xcolor} %used for font color 
\usepackage{float}
\usepackage[version=4]{mhchem}
\usepackage{caption}
\newcommand{\isotope}[2]{\ce{^{#2}#1}}

\begin{document}

\preprint{APS/123-QED}

\title{Cosmogenic activation in detector materials at shallow depths}  % Force line breaks with \\

% \author{
% Sagar S. Poudel,$^{1}$
% \email{sagar.sharmapoudel@sdsmt.edu}
% Lekhraj Pandey,$^{2}$
% \email{lekhraj.pandey@coyotes.usd.edu}
% Robert Calkins,$^{3}$ Manish K. Jha,$^{2}$ Ben Loer,$^{4}$ 
% John L. Orrell,$^{4}$ \\Alan Robinson,$^{5}$ Joel Sander,$^{2}$ and Richard W. Schnee$^{1}$
% }

% \affiliation{
% $^{1}$Department of Physics, South Dakota School of Mines and Technology, Rapid City, SD 57701, USA \\
% $^{2}$ Department of Physics, University of South Dakota, Vermillion, SD 57069, USA \\
% $^{3}$Department of Physics, Southern Methodist University, Dallas, TX 75275, USA
% \\
% $^{4}$Pacific Northwest National Laboratory, Richland, WA 99352, USA \\
% $^{5}$D\'epartement de Physique, Universit\'e de Montr\'eal, Montr\'eal, Québec H3C 3J7, Canada
% }
\author{Sagar S. Poudel}
\email{sagar.sharmapoudel@sdsmt.edu}
\affiliation{Department of Physics, South Dakota School of Mines and Technology, Rapid City, SD 57701, USA}

\author{Lekhraj Pandey}
\email{lekhraj.pandey@coyotes.usd.edu}
\affiliation{Department of Physics, University of South Dakota, Vermillion, SD 57069, USA}

\author{Robert Calkins}
\affiliation{Department of Physics, Southern Methodist University, Dallas, TX 75275, USA}

\author{Manish K. Jha}
\affiliation{Department of Physics, University of South Dakota, Vermillion, SD 57069, USA} 

\author{\mbox{Ben Loer}}
\affiliation{Pacific Northwest National Laboratory, Richland, WA 99352, USA}

\author{John L. Orrell}
\affiliation{Pacific Northwest National Laboratory, Richland, WA 99352, USA}

\author{Alan Robinson}
\affiliation{D\'epartement de Physique, Universit\'e de Montr\'eal, Montr\'eal, Qu\'ebec H3C 3J7, Canada}

\author{Joel Sander}
\affiliation{Department of Physics, University of South Dakota, Vermillion, SD 57069, USA}

\author{Richard W. Schnee}
\affiliation{Department of Physics, South Dakota School of Mines and Technology, Rapid City, SD 57701, USA}

\smallskip

\begin{abstract}
The radioactive decay from long-lived radioactive isotopes produced by cosmogenic activation can be an important background in direct-detection dark matter and neutrino experiments. In general, activation of materials located above ground is dominated by nuclear spallation due to energetic neutrons produced as secondary particles from primary cosmic ray interactions in the atmosphere. As experiments become larger and strive for greater sensitivity to rare events, it is increasingly important to store, assemble, and even fabricate the detector materials underground to mitigate cosmogenic activation. There has been no study of cosmogenic activation in detector materials at shallow depths ($<$ 100 meter-water-equivalent). Unlike at aboveground or at deep depths, where neutrons are the major contributors to activation in materials, there are multiple competing physical processes that contribute to the activation in materials at shallow depths. We present a detailed calculation of the production of tritium in Ge and Si, as well as the production of \isotope{Co}{60} in Cu, at shallow depths. We also obtain cosmogenic activation suppression factors and tritium production at several shallow-depth sites including the Stanford Underground Facility (SUF), where the SuperCDMS collaboration stored Ge, Si, and Cu detector materials for a substantial period of time.

\end{abstract}

\maketitle

\section{Introduction}
The radioactivity produced in detector and shielding materials due to their exposure to cosmic ray secondaries can be problematic for experiments carrying out rare-event searches. Extended exposure of shielding and detector materials to cosmic ray secondaries aboveground during storage, transport, and assembly can produce problematic levels of long-lived radioactive isotopes including tritium ($\tau_{1/2}=12.32$~y, $Q=18.2$~keV) in Ge and Si. Tritium decays can contribute to the background in experiments like SuperCDMS \cite{agnese2017projected}, DAMIC \cite{settimo2018damic}, and OSCURA \cite{aguilar2022oscura} which search for low-mass dark-matter interactions using Ge and/or Si detectors. In addition, cosmogenic activation of copper materials, which are used in the cryostat or shield, can also contribute to the background. In particular, the long-lived isotope \isotope{Co}{60} ($\tau_{1/2}=5.3$~y), which is cosmogenically produced in copper, emits highly-penetrating MeV-energy-scale gamma rays. 
In general, as rare-event search experiments become larger, unmitigated cosmogenic activation in detector materials have greater potential to result in background events and limit an experiment's sensitivity.

As a concrete example, the SuperCDMS SNOLAB experiment will use Ge and Si detectors to search for low-mass dark matter interactions. Near a detector's energy sensitivity threshold, discrimination between electron-recoil (ER) signals and nuclear-recoil (NR) signals is less robust. In particular, the SuperCDMS SNOLAB Ge and Si high-voltage (HV) detectors are optimized for the low-mass dark matter search, and lack ER/NR event discrimination \cite{agnese2017projected,albakry2022investigating}. As a result, the SuperCDMS Collaboration made significant efforts to control tritium production in the detectors as well as activation in Cu components. This included \textbf{(i)} underground storage of the detector crystal materials at HADES lab~\cite{2023GSLSP.536} ($\sim$500 meter-water-equivalent (mwe)\cite{andreotti2010status}) when not in use during the manufacture of Ge crystals at Umicore (Umicore Electro-Optic Materials, Olen, Belgium) and Si crystals at TopSiL (Topsil GlobalWafers A/S, Frederikssund, Denmark), \textbf{(ii)} overseas and road transport of Ge and Si crystals in the GERDA/MAJORANA shielded container (suppressing tritium production by a factor of $\sim$10)~\cite{barabanov2006cosmogenic}, \textbf{(iii)} long-term storage of detector (crystals) and Cu components at the Stanford Underground Facility (SUF) in tunnels A and C (15-20~mwe depth), and \textbf{(iv)} storage of SuperCDMS SNOLAB detector towers at the Stanford Linear Collider (SLC) Adit Storage (50-60~mwe depth).
 
While there has been significant study \cite{laubenstein2009cosmogenic,baudis2015cosmogenic,zhang2016cosmogenic,armengaud2017measurement, amare2018cosmogenic,agnese2019production, elersich2023study, ma2019study,she2021study} of the cosmogenic activation in common detector and shielding materials aboveground, and some study \cite{wiesinger2018virtual} at deep underground sites where rare-event search experiments are usually located, there is no equivalent study for cosmogenic activation at shallow depth locations. This report is focused on activation of Ge, Si, and Cu due to exposure to cosmic ray secondaries at shallow depths. As rare-event search experiments get larger and more sensitive, we expect the use of shallow-depth sites to become increasingly common. In addition, the use of shallow underground facilities for crystal growth (as suggested in \cite{mei2024enhancing,platt2018supercdms}) and fabrication of detector materials such as the production of electroformed copper at Pacific Northwest National Laboratory Shallow Underground Lab (PNNL SUL) \cite{hoppe2014reduction}, and for detector assembly makes the study of cosmogenic activation at shallow depth important.

In this work, we develop a cosmogenic activation model to estimate \textbf{(1)} the tritium activity induced in Ge and Si, \isotope{Co}{60} production in Cu at specific shallow depth locations, and \textbf{(2)} more generally provide the tritium and \isotope{Co}{60} production rates and suppression factors at shallow depths.

This study primarily focuses on shallow-depth sites, namely SUF Tunnel A/C, PNNL SUL, and SLC Adit Storage. The mwe depths of these shallow-depth sites are shown in Table~\ref{table:shallow_depth_facility}. \begin{table}[ht!]
\small
\begin{threeparttable}
\begin{tabular}{|p{3.0cm}|p{3.0cm}|}
 \hline
 Facility  & Depth (mwe)\\
 \hline
 \hline
 SUF Tunnel A and C & 15-20\\
 \hline
 PNNL SUL & $\sim 30$ \\
 \hline
 SLC Adit Storage & 50-60\tnote{a}\\
 \hline
\end{tabular}

\begin{tablenotes}
\footnotesize
\item[a] Calculated based on the muon rate measurements made by the SuperCDMS Collaboration.
\end{tablenotes}

\caption{The mwe depth of shallow-site facilities of interest in this work.}
\label{table:shallow_depth_facility}
\end{threeparttable}
\end{table}
Based on ~\cite{chen1993measurements,da1995neutron}, SUF Tunnels should be at 15-20~mwe depth. SUF Tunnel A is a few mwe shallower than the SUF Tunnel C.  For simplicity, all calculations for SUF Tunnels presented in this work assume a fixed depth of 20~mwe. The depth of the PNNL SUL facility is approximately 30~mwe~\cite{aalseth2012shallow}. The mwe depth of the SLC Adit Storage was measured by the SuperCDMS Collaboration using muon witness detector~\cite{aguayo2011cosmic,aguayo2013spl}, and was graciously provided to us for use in this shallow-depth site assessment effort. A factor of 3 smaller muon event rate at the SLC Adit Storage corresponds to about 54 mwe (for the simulated rock, we consider this depth to be 20 m). The mwe depths of SUF Tunnels, PNNL SUL, and  SLC Adit Storage correspond to 7 m, 12 m, and 20~m of overburden of assumed rock~\cite{rumble2020crc} (density of 2.7 g/cm$^3$). The PNNL SUL and SLC Adit storage sites do not have flat overburden; however, the uncertainties in our results for these sites are large enough that the non-flatness of the overburden is a negligible effect.

These sites are relevant for the following reasons: SUF Tunnel A/C, where the SuperCDMS SNOLAB Ge and Si crystals/detectors were stored for 6–8 years. In addition, SUF Tunnel C was also used to store the Cu components that were used in the SuperCDMS SNOLAB housing and detector towers for about 6 years, before they were brought to the surface for detector and tower assembly. In addition, we consider the SLC Adit storage, that is at greater depth and is the underground location used by the SuperCDMS Collaboration to store the detector towers (following assembly) for about a year. We also consider the PNNL SUL, which produces low-radioactivity electroformed copper for rare-search experiments.

The organization of this paper is as follows. In Section \ref{Suppression of cosmic ray secondary neutrons at shallow depths}, we discuss the suppression of cosmic ray secondary neutrons at shallow-depth sites, and the processes that are relevant for our activation study. In Section~\ref{sec:neutron_flux}, we discuss our simulations of neutron flux and energy spectrum at shallow depths which is essential to obtain neutron-induced activation. Neutron-induced activation rates are presented in Section~\ref{Muon-induced neutron cosmogenic activation}. At shallow depths, the stopping rate of muons is high, and activation from stopping negative muon is important. In Section \ref{Stopping_muons_and_cosmogenic_activation}, we discuss the stopping muons and the cosmogenic activation from negative muon capture. Despite the cross-section for real-photon induced interactions being small, the flux of the gammas produced by cosmic-ray muon-induced interactions is high. We discuss photonuclear activation separately in Section \ref{Photo-nuclear activation}. In Section~\ref{Other processes}, we discuss other secondary processes that can produce activation but their contribution is sub-dominant. Finally, in Section~\ref{Result_and_Discuss}, we summarize our findings and discuss the cosmogenic suppression factors for shallow depths obtained in this work. Various auxiliary physics studies, that were carried out to compare our findings with literature, and understand the systematics in the production rates, are presented in the Appendix A-K.

The results presented in this work are valid for materials whose thicknesses are comparable or smaller than the mean free path of $>$~10 MeV neutrons and gammas and where the stopping muon rate can be assumed to be  uniform, which is the case for current generation Ge, Si and Cu materials used in rare-event searches. When detectors are stacked together or the thickness of the material is large, neutron-induced isotope production is enhanced \cite{agafonova2013universal} as the thickness of the material may be enough to allow the hadronic shower development within the material itself, while isotope production induced by stopping negative muons is suppressed due to self-shielding effects.

\section{Suppression of cosmic ray secondary neutrons at shallow depths}
\label{Suppression of cosmic ray secondary neutrons at shallow depths}

At sea-level, cosmogenic activation is primarily caused by the neutrons \cite{ziegler1998terrestrial,cebrian2017cosmogenic} that originate in the upper atmosphere from showers generated by the cosmic-ray interactions with the nuclei of nitrogen and oxygen \cite{gaisser2016cosmic}. The neutron spectra at sea-level obtained using Gordon's parameterization \cite{gordon2004measurement} and EXPACs \cite{sato2015analytical,expacs} is shown in Appendix \ref{sea_level_neutrons}. Storage and fabrication activities at the underground locations studied in this report were intended to significantly reduce the flux of secondary cosmic ray neutrons.  In addition, the flux of the primary cosmic-ray protons, which is significantly reduced at sea-level and that of  pions generated in the upper atmosphere are suppressed completely by approximately 4 m of rock \cite{smith1989naturally}. Our simulation results, described in  Appendix \ref{neutron_flux_attenuation}, show that with 5 m (about 14 mwe) of rock overburden, the flux of $>$ 10 MeV secondary cosmic ray neutrons is an order of magnitude smaller than from the neutrons originating from muon-induced interactions. This is also anticipated by a prior study \cite{vanhoefer2015neutron}, which showed, at 4 m depth for LNGS rock, neutrons from muon-induced interactions is a major contributor to $>$~20 MeV neutron flux. At shallow-depth sites of interest in this study, cosmic-ray muons and the particles (including neutrons and gammas) produced from their interactions will dominate the relevant activation processes.

\section{Muon and neutron flux and energy spectrum at shallow depths}
\label{sec:neutron_flux} 

For the shallow-depth sites of interest in this work, only the neutrons produced from cosmic-ray muon-induced interactions are relevant. In addition, the length of overlying rock overburden is large enough (10 mwe or greater) for full hadronic shower development in rock from muon-induced interactions. For the isotopes of interest in the work, neutron-induced production falls precipitously below approximately 10 MeV. Therefore, neutrons of radiogenic origin (from spontaneous fission and ($\alpha$, n) interactions) are not relevant, as their energy spectrum drops off after a few MeV. Thus, the activation from neutrons of radiogenic origin was not considered (except for calculations of negligible production of \isotope{Co}{60} in Cu) in this work. 

We use a more detailed rock composition (taken from \cite{rumble2020crc}) and shown in Table \ref{table: rock_composition}) in our simulations, representative of the average composition of the Earth's crust. The mass-fraction-weighted effective Z and A for this composition are 12 and 24 (in comparison for standard rock, they are 11 and 22, respectively). Neutrons produced through negative muon capture neutrons depend on the rock composition motivating using a more realistic rock composition. 

Variations in the rock composition from the average composition assumed in this study will affect the $>$ 10 MeV neutron flux at the few-percent level (the simulation results for crust and limestone compositions shown in Appendix \ref{crust_vs_limestone} also demonstrate this). This is also expected because hadronic shower neutron production is weakly dependent \cite{agafonova2013universal} on composition ($\propto$ A$^{0.95}$, where A is the mass number and can be considered mass-fraction-weighted for composite materials). Those neutrons dominate the production rate compared to negative-muon-capture neutrons given the high energy threshold for neutron induced activation considered in this study.

Using different rock compositions is expected to change the neutron-induced activation rates by a few percent for tritium production in Ge and Si, but a few tens of percent for \isotope{Co}{60} in Cu. 

\begin{table}[ht!]
 \small
\begin{tabular}{|p{3.0cm}|p{3.0cm}|}
 %\multicolumn{5}{}{Major $\beta$-decay channels in Ar42 chain}\\ 
 \hline
 Elements  & Mass fraction (in $\%$)\\
 \hline
 \hline
 Oxygen (O) & 46.10\\
 \hline
 Silicon (Si) & 28.20\\
 \hline
 Aluminium (Al) & 8.23 \\
 \hline
 Iron (Fe) & 5.63\\
 \hline
 Calcium (Ca) & 4.15\\
 \hline
 Sodium (Na) & 2.36\\
 \hline
 Magnesium (Mg) & 2.33\\
 \hline
 Potassium (K) & 2.09\\
 \hline
 \hline
 
 \end{tabular}
  \caption{Elemental abundances in the modeled crust composition used in this work. The rock composition is taken from \cite{rumble2020crc} for upper continental crust. The rock density is taken to be 2.7 g/cm$^3$.}
  \label{table: rock_composition}
\end{table}
In our simulations, we generate cosmic-ray showers specific to New York City (NYC), 2003, using the Cosmic-Ray Shower generator (CRY) \cite{hagmann2007cosmic,hagmann2012cosmic}. Choosing that reference sea-level elevation is motivated by Gordon et al.’s measurements \cite{gordon2004measurement} in NYC in 2003, for which reference neutron flux and activation rates are available for comparison with our results.  The resulting normalization to obtain rates is based on the CRY-given muon flux of 1.15 $\times$ 10$^{-2}$ muons / (cm$^{2}$ s). 

The description of the simulation of cosmic-ray muon interactions, and the recording of the resulting muons and neutrons, is as follows. We generate cosmic ray muons in a 100 m $\times$ 100 m plane, a plane that is large enough for the shallow-depth sites of interest, using CRY and use FLUKA \cite{FLUKA_1_battistoni2015overview, FLUKA_2_bohlen2014fluka, FLUKA_3_ballarini2007physics} simulations to propagate muons and muon-induced secondary particles in the rock.  We record the muon-induced neutron spectrum and the flux at various depths in the rock. We turn on all the physics processes and transport the secondary particles that are relevant for neutron production. Details are in Appendix \ref{FLUKA simulations and particle transport}. In the simulations, the negative muons are transported all the way to keV energy so as to allow for negative muon capture and production of muon-capture neutrons. Neutrons with energies of 1 MeV or lower aren't transported in the simulations as they are not able to to produce the activation relevant for our study as the threshold for neutron induced activation reactions are much higher.

\begin{figure}[h]
\centering
\includegraphics[width=1.0\linewidth]{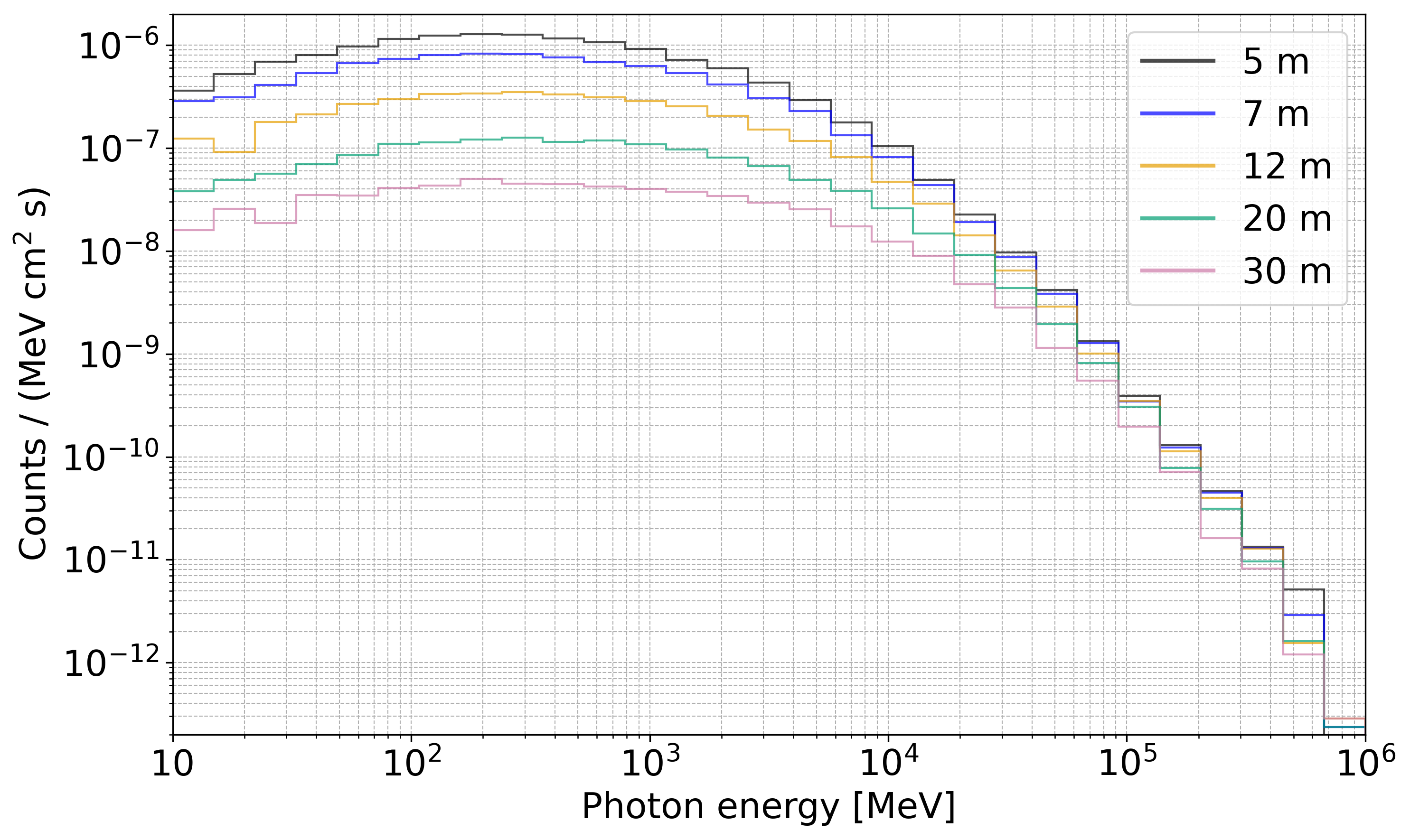}
\caption{The simulated muon flux and spectrum at various depths in the crust of the Earth.}
\label{Figure: muon_spectrum}
\end{figure}

\begin{figure}[h]
\centering
\includegraphics[width=1.0\linewidth]{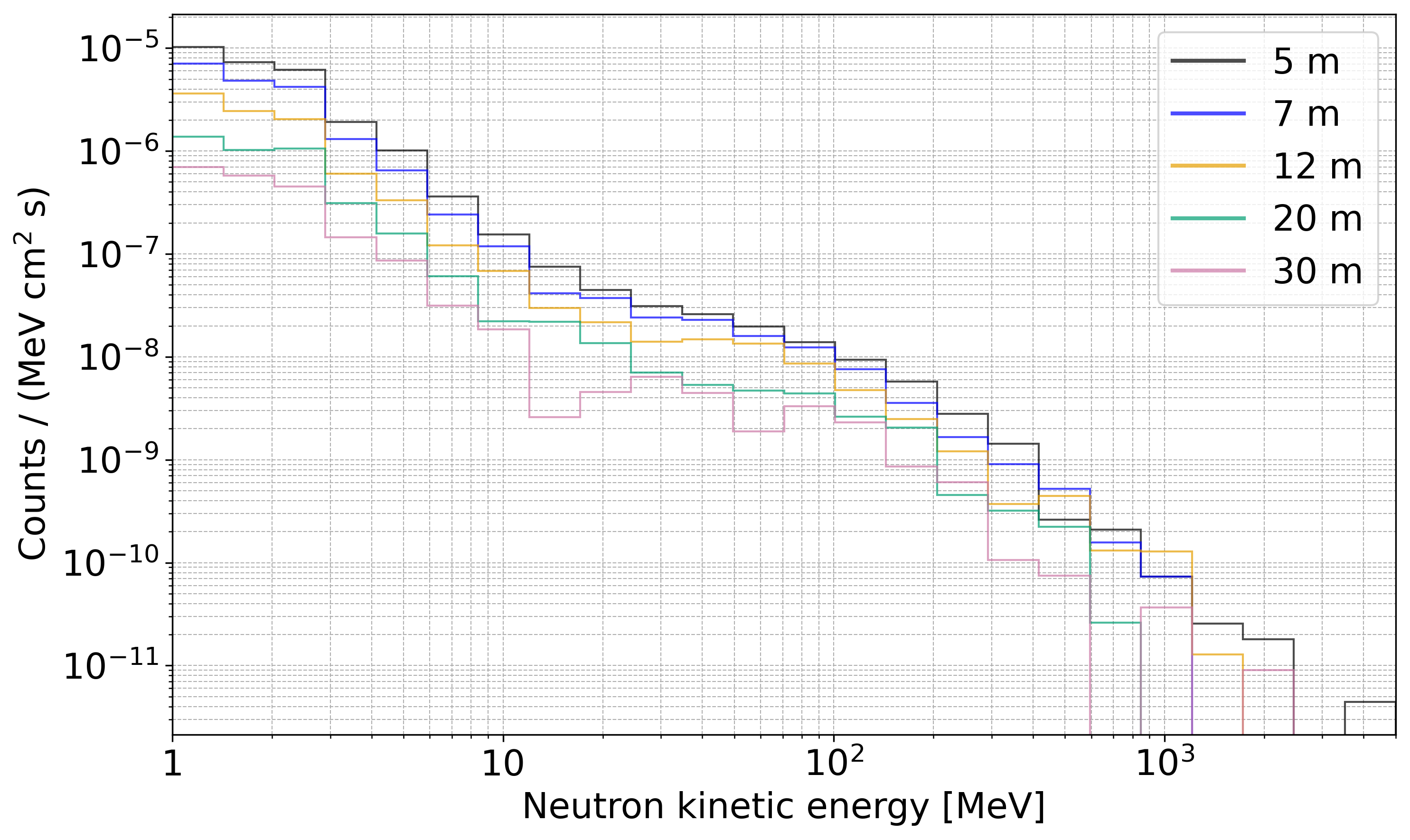}
\caption{The simulated muon-induced neutron flux and spectrum at various depths in the crust of the Earth.}
\label{Figure: neutron_spectrum}
\end{figure}

Simulated neutrons are recorded when they cross 10 m x 10 m planes located at depths of at 5 m, 7 m, 12 m, 20 m and 30 m. The depths 7 m, 12 m, and 20 m, correspond approximately to the mwe depths of SUF Tunnel A/C, PNNL SUL, and SLC Adit Storage, respectively. The simulated muon and neutron fluxes spectrum obtained from the simulations are shown in Figures \ref{Figure: muon_spectrum} and \ref{Figure: neutron_spectrum}. The muon and $>$ 10 MeV neutron flux  are reported in the Table \ref{table: Muon-neutron-flux}. CRY-given muon flux and $>$ 10 MeV neutron flux (from \cite{gordon2004measurement} ) at sea level are also listed in the table for comparison. These results are based on 1.74 $\times$ 10$^8$ muons generated in a 100 m x 100 m plane. The statistical uncertainty in the muon flux is negligible while in the $>$ 10 MeV neutron flux is 10$\%$ or lower. 

\begin{table}[ht!]
 \small
\begin{tabular}{|p{1.0cm}|p{3.2cm}|p{2.0cm}|p{2.1cm}| }
 \hline
 \textbf{Depth} (m)& \textbf{Description} & \textbf{Muon flux (/cm$^{2}$/s)} & \textbf{$>$ 10 MeV neutron flux (/cm$^2$/s)} \\
 \hline
 0  & sea-level & 1.15 $\times$ 10$^{-2}$ & 3.5 $\times$ 10$^{-3}$ \\
 \hline
 5  & muon-induced $>$ 10 MeV neutrons start dominating &5.10 $\times$ 10$^{-3}$ & 3.9 $\times$ 10$^{-6}$ \\
 \hline
 7  & SUF Tunnels A/C  & 3.80 $\times$ 10$^{-3}$ & 3.0 $\times$ 10$^{-6}$ \\
 \hline
 12 & PNNL SUL  & 2.08 $\times$ 10 $^{-3}$ & 2.0 $\times$ 10$^{-6}$ \\
 \hline
 20 & SLC Adit Storage & 9.94 $\times$ 10$^{-4}$ & 1.0 $\times$ 10$^{-6}$ \\
 \hline
 \end{tabular}
  \caption{The simulated muon flux and flux of $>$ 10 MeV neutrons from muon-induced interactions.}
  \label{table: Muon-neutron-flux}
\end{table}

Chen et al. (1993) \cite{chen1993measurements} measured the neutron flux at SUF in the energy range (11.5-50) MeV to be 1.07$_{-0.30}^{+0.41}$ ~ $\times$ 10$^{-6}$ n / (cm$^{2}$ s). For our modeled rock, the neutron flux in that energy range is 1.2 $\times$ 10$^{-6}$ n / (cm$^{2}$ s) . This shows our FLUKA simulations reproduce well the measured neutron flux even at shallow depths.
As discussed in Appendix \ref{crust_vs_limestone}, if we assume the same m.w.e. thickness of rock overburden but change the rock composition to limestone, the $>$ 10 MeV neutron flux would be greater by about 10\%. Considering that the simulations reproduce the measurements at SUF well and that the measurements also have about 40\% uncertainty, we conservatively estimate the uncertainty in the $>$ 10 MeV neutron flux to be about 50\%.

\section{Muon-induced neutron cosmogenic activation}
\label{Muon-induced neutron cosmogenic activation}
In this section, we discuss the neutron-induced tritium production in Ge and Si and \isotope{Co}{60} production in copper at shallow depths. A comparison to the sea-level production rate is also presented. 
\subsection{Neutron-induced tritium production in Ge and Si}
\begin{figure}[ht]
\centering
\includegraphics[width=1.0\linewidth]{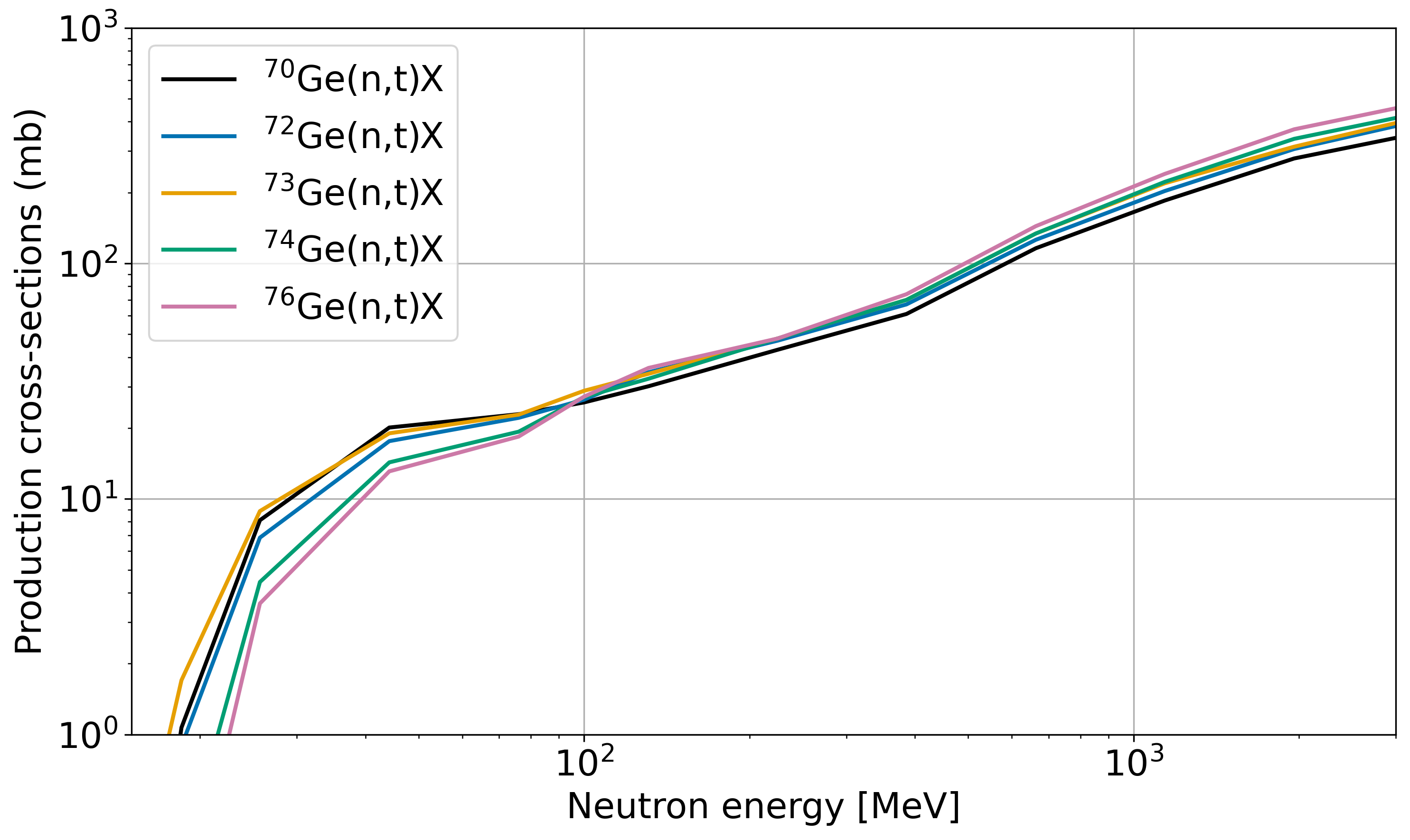}
\caption{Tritium production cross-sections as a function of neutron kinetic energy in Ge.}

\label{Figure: cross_Ge}
\end{figure}

\begin{figure}[h]
\centering
\includegraphics[width=1.0\linewidth]{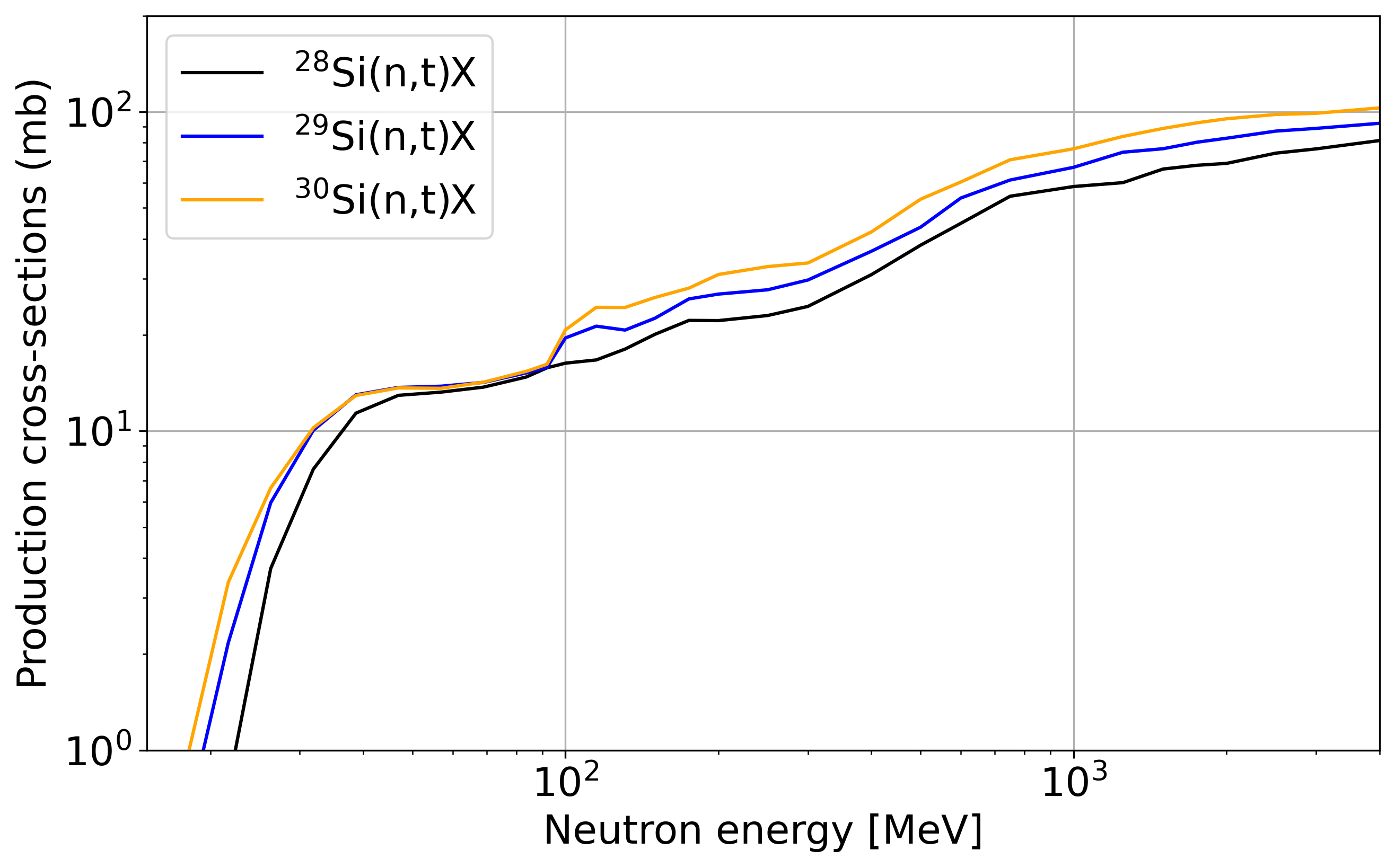}
\caption{Tritium production cross-sections as a function of neutron kinetic energy in Si.} 
\label{Figure: cross_Si}
\end{figure}

To estimate the rate of tritium production from neutrons in Ge and Si, we follow \cite{agnese2019production,saldanha2019cosmogenic} and use a mix of nuclear models: TALYS  \footnote{The use of TALYS alone for the entire energy range gives a tritium production a factor of about 5 times higher at the depth of SUF.} for $<$ 100 MeV neutron nergies and INCL for $\geq$ 100 MeV neutron energies and simulated neutron flux. The production cross-sections are shown in Figures \ref{Figure: cross_Ge} and \ref{Figure: cross_Si}. Using this model for production cross-sections and Gordon's neutron spectrum, we obtain the sea-level tritium production rate of  94 (125) atoms per kg Ge (Si) per day, which agrees within 25$\%$ of the sea-level production rates based on the tritium activation measurements in Ge \cite{agnese2019production,armengaud2017measurement} and Si \cite{saldanha2019cosmogenic}. Using the neutron flux in Figure \ref{Figure: neutron_spectrum}, and the production cross-sections in Figure \ref{Figure: cross_Ge} and \ref{Figure: cross_Si} \, respectively, we obtain tritium production rates considering the natural isotopic elements of corresponding elements \footnote{Ge: \isotope{Ge}{70}: 20.4\%, \isotope{Ge}{72}: 27.3\%, \isotope{Ge}{73}: 7.8\%, \isotope{Ge}{74}: 36.7 \%, \isotope{Ge}{76}: 7.8\% and Si: \isotope{Si}{28}: 92.2\%, \isotope{Si}{29}: 4.7\%, \isotope{Si}{30}: 3.1\% }  in Ge and Si at various depths in rock. 

In an actual cavern at shallow-depths, we expect some contribution to the cosmogenic activation from the neutrons that are back-injected into the cavern. Based on the simulation results reported in Appendix \ref{back-injected neutrons}, we estimate that back-injected neutrons contribute an additional ~40\% to the total neutron-induced activation. Including this enhancement, we obtain the neutron-induced tritium production rates at the SUF Tunnels A/C, PNNL SUL, and SLC Adit Storage sites, as reported in Table \ref{table:production_rate_neutrons}.

\begin{table}[ht!]
\small
\begin{threeparttable}
\begin{tabular}{|p{4.0cm}|p{1.2cm}|p{1.2cm}|p{1.2cm}|}
\hline
\multirow{2}{*}{Location} & Si & Ge & Cu \\ \cline{2-4}
 & \isotope{H}{3} & \isotope{H}{3} & \isotope{Co}{60} \\ 
\hline
Sea-level & 125 & 94 & 46 \\  
\hline
SUF Tunnel A/C (15--20 mwe) & 0.11 & 0.080 & 0.080 \\
\hline
PNNL SUL ( 30 mwe) & 0.075 & 0.055 & 0.052 \\
\hline 
SLC Adit Storage (50--60 mwe) & 0.036 & 0.027 & 0.025 \\
\hline
\end{tabular}
\end{threeparttable}
\caption{Activation rate (atoms per kg per day) from neutrons at shallow-depth sites. Sea-level activation rates are also shown for comparison.}
\label{table:production_rate_neutrons}
\end{table}

Based on the Chen et al. \cite{chen1993measurements}'s study for SUF, muon-capture neutrons contribute 38\%, while hadronic shower neutrons contribute about 62\% to the neutron flux in the range (11.5-50) MeV. The tritium production we obtain in this energy range is about 20$\% $ of the total. Given that the neutron-induced tritium production is dominated by neutrons with energies $>$ 50 MeV, neutrons produced in muon-induced hadronic showers dominate the neutron-induced tritium production even at shallow depths. The neutron yield studies from \cite{barton1985spectrum,malgin2017energy} also shows hadronic showers dominates the  neutron production in rock at energies $>$ 30 MeV.  
\subsection{Neutron-induced \isotope{Co}{60} production in copper}
\label{60Co-production}
At sea-level, cosmic-ray neutron-induced \isotope{Co}{60} production in copper dominates over other processes. There are three reported measurements of sea-level \isotope{Co}{60} production rates (in atoms per kg Cu per day):  
$39.7 \pm 5.7$ (She et al. \cite{she2021study}), $29.4^{+7.1}_{-5.9}$ (Baudis et al.\cite{baudis2015cosmogenic}), and $86.2 \pm 7.6$ (Laubenstein et al.\cite{laubenstein2009cosmogenic}).

\begin{figure}[h]
\centering  
\includegraphics[width=1.0\linewidth]{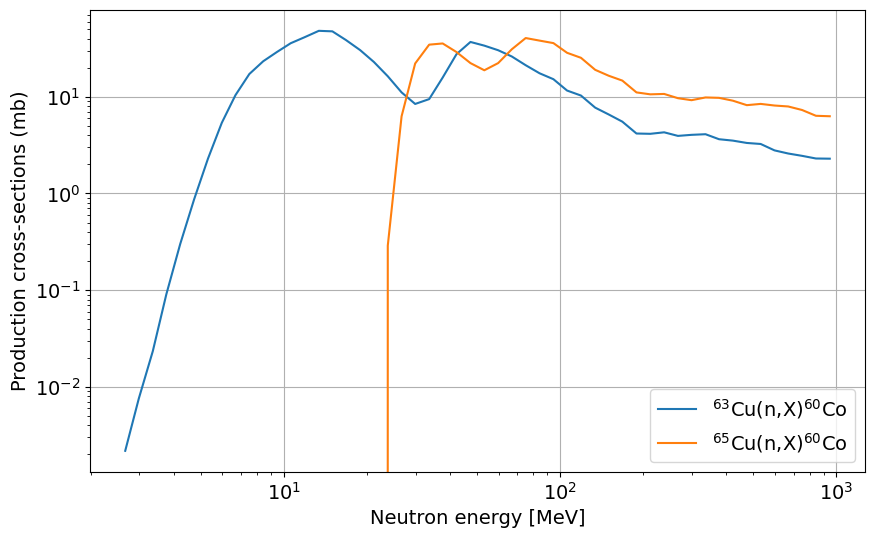}
\caption{\isotope{Co}{60} production cross-sections as a function of neutron kinetic energy in copper. The production rates from \isotope{Cu}{63} (shown in black) and \isotope{Cu}{65} (shown in blue) are shown separately.} 
\label{Figure: cross_Cu}
\end{figure}

 In comparison, one gets a sea-level production rate of \isotope{Co}{60} 53 \footnote{ Gordon’s measured data are used instead of his parameterization, so the contribution from neutrons below 10 MeV is included}  (47) per kg of copper per day using TALYS \cite{TALYS_1goriely2008improved,TALYS_2koning2012modern}  cross-sections (shown in Figure \ref{Figure: cross_Cu}) and Gordon's (EXPACS \cite{sato2015analytical}) neutron spectrum. We consider Susan 2017 \cite{cebrian2017cosmogenic}'s production rate of 46 \isotope{Co}{60} atoms per kg of Cu per day (based on TALYS cross-sections and Gordon's neutron spectrum) as a reference sea-level production rate to compare the production rates to that of shallow depths. This production rate agrees well within 50$\%$ to all reported measurements. This is why we use TALYS cross-sections to estimate \isotope{Co}{60} production rates for shallow-depth sites. Using the cross-sections  and the neutron spectra (in Figure \ref{Figure: cross_Cu}), obtain \isotope{Co}{60} production rates at various depths. The total production rate is obtained by summing the production rate from both channels weighing by the natural isotopic abundances of \isotope{Cu}{63} (69.17$\%$) and \isotope{Cu}{65} (30.83$\%$). An additional 40$\%$ production is included to consider the contribution from back-injected neutrons reported in Appendix \ref{back-injected neutrons}. The results for various shallow-depth sites are shown in Table \ref{table:production_rate_neutrons}.

As seen in the cross-section plot in Figure \ref{Figure: cross_Cu}, even though the \isotope{Co}{60} production cross-section peaks above 10 MeV neutron energy, the reaction \isotope{Cu}{63}(n,X)\isotope{Co}{60} has an energy threshold of less 10 MeV, and the production from such low-energy neutrons needs to be considered.   
Using the Gordon's sea-level measured neutron spectrum shown in Figure \ref{Figure: Neutron_comparison}, the relative contribution of $<$ 10 MeV neutrons to the \isotope{Co}{60} production at sea level is about 5$\%$. In comparison, we find that at the depth of SUF, the contribution from $<$ 10 MeV neutrons is about 30$\%$ (while it is 10$\%$ at the depth of SLC Adit storage). The reason for this is the less energetic spectrum of muon-induced neutrons at shallow depths. From the comparison shown in Appendix \ref{back-injected neutrons}, the mean energy of $>$ 10 MeV neutrons at the depth of SUF (20 mwe) is about 90 MeV while 170 MeV at sea-level. Neutrons from negative muon capture \cite{suzuki1987total} tend to populate the low-energy part of the neutron flux spectrum, increasing its softness. In contrast, the $>$ 10 MeV neutron spectrum at sea level is expected to be hard, since the bulk of energetic sea-level neutrons originate in the upper atmosphere and low-energy neutrons are attenuated as they traverse the atmosphere. Based on the study of Chen et al. 1993,  muon-capture neutrons contribute about 38$\%$ to (11.5-50) MeV neutron flux at the depth of SUF. From our calculation, neutrons in that energy range produce about 60\% of the total neutron-induced \isotope{Co}{60} (in Cu) at SUF (and about 50$\%$ at SLC Adit Storage). In comparison, the corresponding number for tritium production in Ge and Si at these depths is about 20\%. This suggests muon-capture neutrons are an important contributor to the neutron-induced  \isotope{Co}{60} production in copper at shallow depths. Since muon-capture neutron flux depends more strongly on the  composition of the rock overburden, it is useful to gain knowledge of the composition of the rock overburden in the case where copper materials are intended to be stored. Based on our study (shown in Appendix \ref{crust_vs_limestone}) of $>$ 10 MeV neutron flux and spectrum in crust composition (used in this work) and limestone (where Ca content is a factor of 10 higher) , neutron-induced \isotope{Co}{60} production in copper could vary about 30$\%$ depending on the composition of rock. For tritium production, which peaks at much higher energies, the difference is smaller.

\section{Stopping muons and cosmogenic activation}
\label{Stopping_muons_and_cosmogenic_activation}
At shallow depths, the activation from capture of stopping negative muons is important because the stopping muon rate is high.  In this section, we present our calculation of tritium production in Ge and Si, and \isotope{Co}{60} production in copper at shallow depths from capture of stopping negative muons. The isotope production rate ( in atoms per kg per day) from stopping muons is given by:
\begin{equation}
P(X) = R_0 \cdot \left( \frac{\lambda_c}{Q \lambda_d + \lambda_c} \right) \cdot f(X),
\label{eq:1}
\end{equation}
Here, $R_0$ is the negative muon stopping rate in material (which is largely independent of the material). The term in the parentheses can be interpreted as the negative muon capture probability which depends on the material type, and $f(X)$ is the probability of emission of a particle following negative muon capture. $\lambda_c$ and $\lambda_d$ are negative muon capture and muon decay rate, while $Q$ is Huff's factor \cite{huff1961decay} which accounts for the fact that the decay rate of muons when bound to an atomic orbital is slightly suppressed compared to when muons decay into free space.
With the $\lambda_c$ and $Q$ from Suzuki et al. 1987 \cite{suzuki1987total}, we obtain negative muon capture probability for  Si, Ge, and Cu to be  0.65, 0.93, and 0.93 (uncertainties on these numbers are less than 10$\%$) respectively. These capture probabilities are used in Equation \ref{eq:1} to determine the activation rates from stopping muons.
\subsection{Stopping negative muon rate}
\label{Stopping negative muon rate}
We use S. Charambulus's parameterization \cite{charalambus1971nuclear} of negative muon stopping rate as a function of depth for the shallow depth of interest. Negative muon stopping rates for the sites listed in Table~\ref{table:shallow_depth_facility} are shown in the Table \ref{table:muon-stopping-rate}. The negative muon stopping rate at sea level, based on the measurement \cite{barton1965intensity} taken at a geomagnetic latitude of $40^\circ$ \cite{charalambus1971nuclear}, is also listed for comparison. 

 \begin{table}[ht!]
\small
\begin{tabular}{|p{3.0cm}|p{3.0cm}|}
 \hline
 Location   & Negative muon stopping rate (/kg/day) \\
 \hline
 Sea-level  & 483 $\pm$ 52\\ 
 \hline
 SUF Tunnel A/C (15--20 mwe) & 73 \newline 90 $\pm$ 11*\\
 \hline
 PNNL SUL (30 mwe)  & 34\\
 \hline
 SLC Adit Storage (50--60 mwe) & 13\\
 \hline
\end{tabular}
\caption{Muon stopping rate at some shallow-depth sites obtained from parameterization in \cite{charalambus1971nuclear}. In * is the negative muon stopping rate obtained for SUF Tunnel C by propagating the CRY-generated aboveground muons through 20 m.w.e. of modeled rock, as determined in this work from simulations.}
\label{table:muon-stopping-rate}
\end{table}
We also obtain the negative muon stopping rate at SUF Tunnel C propagating the CRY-generated negative muons through 20 mwe of rock. The details are discussed in Appendix \ref{stopping_muon_at_SUF} and the rate is listed in Table \ref{table:muon-stopping-rate} alongside the one from Charambulus's parameterization. Based on the study in \cite{musy2023reviewing}, various parameterizations of negative muon stopping rate as a function of depth agree within 35\% at shallow depths. 

Since at shallow depths the stopping muon rate falls very fast with increasing depth, a lack of precise knowledge of mwe depth also contributes to some uncertainty. For all sites listed in Table \ref{table:muon-stopping-rate}, the uncertainty in the muon stopping rate is expected to be about 50$\%$. We expect the systematic uncertainty arising from variations in the low-energy muon flux—which contributes to the uncertainty in the aboveground negative stopping-muon rate—to be subdominant at shallow depths greater than 15 mwe, since muons with energies below a few GeV would stop in rock and therefore not reach those depths.

\subsection{Production yield per stopped negative muon}
\label{Production yield per stopped negative muon}
n this section, we discuss tritium yield in Ge and Si and \isotope{Co}{60} yield in copper from the capture of negative muons, and we also present existing measurements and references in the literature, along with the yields obtained using FLUKA simulations.
There are a few measurements of the tritium yield in Si \cite{budyashov1971charged, edmonds2022measurement} from negative muon capture but none for tritium yield in Ge or \isotope{Co}{60} yield in Cu. 

The ALCAP collaboration measured the tritium yield \cite{edmonds2022measurement} from negative muon capture in Si by measuring the tritium in a kinetic energy range of 6-17 MeV. The yield is $(1.70 \pm 0.08_{\mathrm{stat}} \pm 0.10_{\mathrm{sys}})\times 10^{-3}$ per negative muon capture. Minato et al. \cite{minato2023nuclear}, present various nuclear models for tritium yield in \isotope{Si}{28} following negative muon capture. Among the models, the STDA + MEC (SGII) model gives a tritium yield of 1.63 $\times$ 10$^{-3}$ per negative muon capture in the energy range of 6-17 MeV on \isotope{Si}{28}, which agrees best with ALCAP's measurement. Taking STDA + MEC (SGII) model to be an accurate representation of the capture rate, we use the total tritium yield per negative muon capture reported in the paper, which is 2.16 $\times$ 10 $^{-3}$ per negative muon capture. Since the bulk of the natural Si is made up of \isotope{Si}{28} (92$\%$ isotopic abundance), we neglect isotopic effects on the yields for various Si isotopes and use this as the tritium yield for natural silicon. Using the negative-muon-capture probability of 0.65 \cite{suzuki1987total}, we obtain the tritium yield per stopped negative muon of 3.35 $\times$ 10$^{-3}$. Saldanha et al. \cite{saldanha2020cosmogenic} obtained the same yield for Si from their calculation, which used the tritium yield spectrum measured by ALCAP in Al \cite{gaponenko2020charged}, but scaled it to match the tritium yield measurement (for energies $>$ 24 MeV) by Wyttenbach et al. \cite{wyttenbach1978probabilities}. 

Since there are no measurements of yields in Ge and \isotope{Co}{60} , we use the yield per stopped negative muon obtained from FLUKA simulations. The yields are tabulated in Table \ref{table: yield_per_stopped_muons}. The kinetic spectra of the tritium following negative muon capture obtained from FLUKA simulations for Ge and Si are shown in Figure \ref{Figure: Kinetic_spectrum_tritium} in Appendix  \ref{Particle_yield_from_stopped_muons_FLUKA}. Even though the total tritium yield for Si obtained using FLUKA simulations agrees well with the STDA + MEC (SGII) model, comparing the yield in the energy range (6–17 MeV), in which ALCAP measured the tritium yield in silicon, shows that FLUKA underestimates the yield by about 30\%, suggesting a discrepancy in the shape of the spectrum produced by FLUKA compared with the measurements.

\begin{table}[ht!]
\small
\begin{tabular}{|p{2.7cm}|p{2.7cm}|p{2.7cm}|}
\hline
Tritium yield in Si & Tritium yield in Ge & \isotope{Co}{60} yield in Cu\\
\hline
3.35 $\times$ 10$^{-3}$  & 1.26 $\times$ 10$^{-3}$ & 4.10 $\times$ 10$^{-3}$ \\
\hline
\end{tabular}
\caption{Yield per stopped negative muon obtained using FLUKA simulations. Statistical uncertainties are $<$ 2\%.}
\label{table: yield_per_stopped_muons}
\end{table}

Since we use a model that reproduces the tritium yield measured by ALCAP for Si, the uncertainty in the yield per negative muon capture is $<$ 25\%, taking into account the uncertainties in both the measured yield and the muon capture probability. Based on the  trendline in \cite{wyttenbach1978probabilities}, the tritium production from negative muon capture in Ge should be about a factor of 3 smaller compared to Si. This is consistent with what we obtain from FLUKA. Based on this result, we expect uncertainty in the yield of tritium from negative muon capture in Ge to be about 50$\%$ or less. There are no measurements for Cu; we consider the uncertainty in the yield of 200$\%$. The uncertainty in muon capture probability is negligible (10$\%$ or less) in all Ge, Si, and Cu \cite{suzuki1987total}.

\subsection{Activation from stopping negative muons}
  Based on the negative stopping muon rate and yield per stopped muons, we can calculate tritium production in Ge and Si, and \isotope{Co}{60} production in copper. 
  The activation rates for shallow-depth sites of interest are reported in Table \ref{table: yield_per_stopped_muons}. 

 \begin{table}[ht!]
\small
\begin{threeparttable}
\begin{tabular}{|p{3.0cm}|p{1.5cm}|p{1.5cm}|p{1.5cm}|}
\hline
\multirow{2}{*}{Location} & Si & Ge & Cu \\ \cline{2-4}
 & Tritium & Tritium & \isotope{Co}{60} \\ 
\hline
Sea-level & 1.6 & 0.61 & 2.0 \\  
\hline
SUF Tunnel A/C (15--20 mwe) & 0.24 & 0.092 & 0.30 \\
\hline
PNNL SUL (30 mwe) & 0.11 & 0.043 & 0.14 \\
\hline 
SLC Adit Storage (50--60 mwe) & 0.044 & 0.016 & 0.053 \\
\hline
\end{tabular}
\end{threeparttable}
\caption{Activation rate (atoms per kg per day) from stopped negative muons at shallow-depth sites. Calculation at sea-level is also given for comparison.}
\label{table:production_rate_stopped_muons}
\end{table}
Due to the much higher tritium yield in Si compared to Ge from negative-muon capture (about a factor of 3), while having similar neutron-induced tritium production cross-sections, neutron-induced tritium production in Si only becomes comparable to that from stopped negative muons at the depth of SLC Adit Storage (50–60 mwe), where the stopped negative muon rate has fallen significantly. In contrast, in germanium, neutron-induced production competes with production from stopped negative muons even at depths such as SUF. Also, the results indicate the neutron-induced \isotope{Co}{60} production in Cu start becoming comparable to that from negative muon-capture at depths greater than $>$ 80 mwe.

\section{Photo-nuclear activation}
\label{Photo-nuclear activation}
While the photo-nuclear activation cross-sections are typically smaller compared to neutron-induced activation, the cosmic-ray muon-induced interactions in rock produce a much higher flux of energetic gammas compared to neutrons. The energetic gammas originate from muon-induced interactions from physical processes: pair production, bremstraahlung, and electromagnetic and hadronic showers. Since the energy required to produce tritium in Ge and Si, and \isotope{Co}{60} from photon interactions, is larger than the energies of gamma rays from natural radioactivity and neutron capture, we consider only gamma rays produced in muon-induced interactions.  We obtain separately the production of tritium in Ge and Si, and production of \isotope{Co}{60} in copper from real photonuclear interactions at shallow depths. 
\subsection{$>$ 10 MeV photon  flux at shallow depths}
\begin{figure}[h]
\centering
\includegraphics[width=1.0\linewidth]{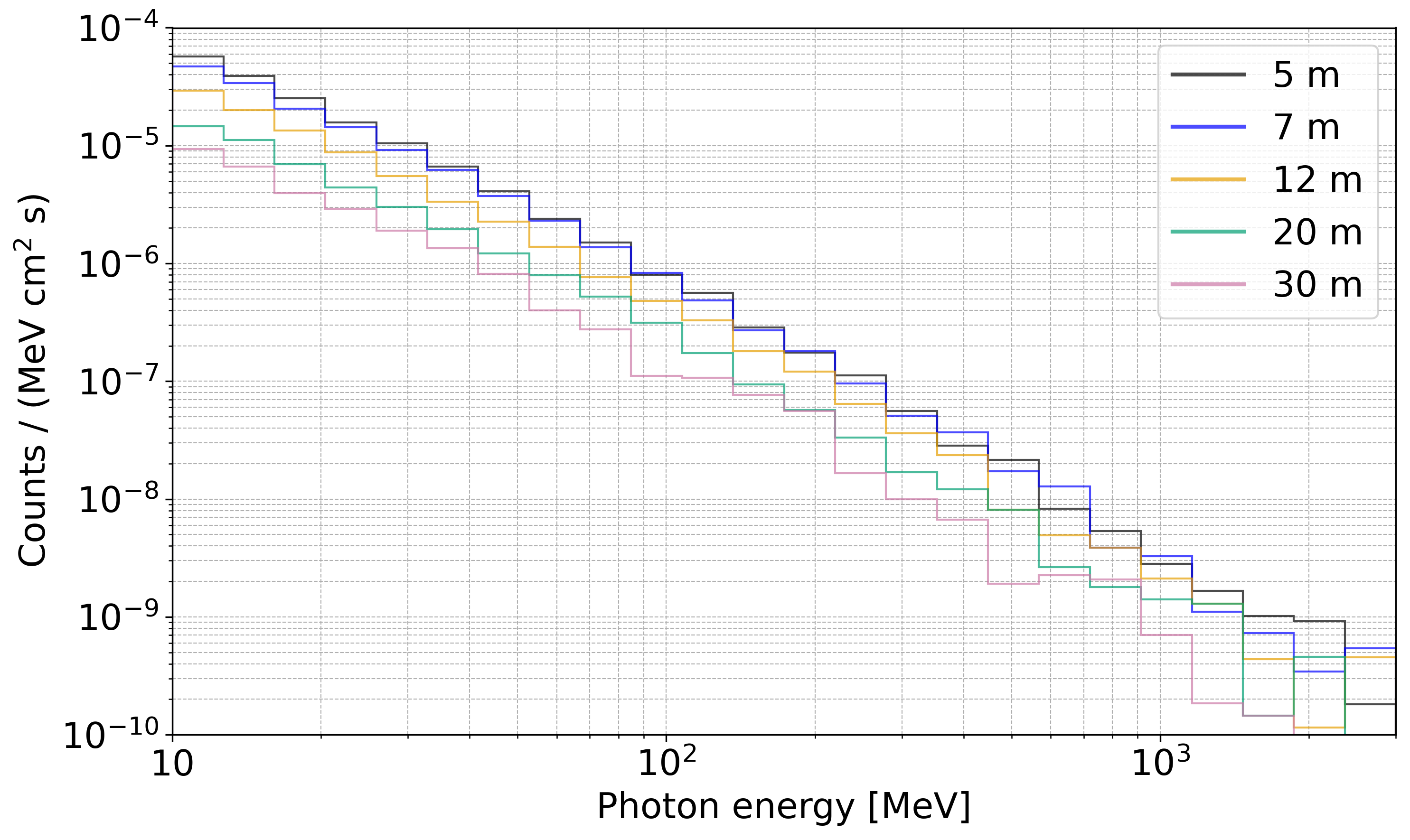}
\caption{Simulated muon-induced gamma flux and spectrum at various depths in the crust}
\label{Figure: gamma_spectrum}
\end{figure}

  In this work, we propagate CRY-code generated muons and muon-induced secondaries and record the flux of $>$ 10 MeV gammas at various depths in rock. The differential (in energy) gamma fluxes for multiple depths in the rock are shown in Figure \ref{Figure: gamma_spectrum}. The expected $>$ 10 MeV gamma flux at various shallow-depth sites of interest obtained from the simulations are presented in Table \ref{table:Muon-gamma-flux}. The sea-level $>$ 10 MeV gamma flux obtained from EXPACS \cite{sato2015analytical} using location NYC, 2003 is also shown in the table  for comparison.

\begin{table}[ht!]
\small
\begin{tabular}{|>{\raggedright\arraybackslash}p{3.5cm}|p{2.1cm}|}
\hline
\textbf{Location} & \textbf{Flux(/cm$^2$/s)} \\
\hline
Sea-level & 9.5 $\times$ 10$^{-3}$ \\
\hline
SUF Tunnels A/C (15--20 mwe)  & 7.0 $\times$ 10$^{-4}$ \\
\hline
PNNL SUL  (30 mwe) & 4.3 $\times$ 10$^{-4}$ \\
\hline
SLC Adit Storage (50--60 mwe) & 2.3 $\times$ 10$^{-4}$ \\
\hline
\end{tabular}
\caption{The $>$10 MeV gamma flux at various shallow-depth sites ( Uncertainties in the numbers are less than 2$\%$). Sea-level gamma flux (obtained using EXPACS for NYC, 2003) is shown for comparison.}
\label{table:Muon-gamma-flux}
\end{table}

The flux of $>$~10~MeV gammas shown in Table \ref{table:Muon-gamma-flux} is about of factor 200 higher than for neutrons. A similar difference is also expected at deeper depths \cite{li2014first} where muon-induced interactions are the source of such gammas. The orders of magnitude difference of flux is expected given the greater abundance of physical processes that can produce energetic gammas from muon-induced interactions.  In comparison, at sea level, the flux of $>$ 10 MeV gamma is just about a factor of 3 higher. This is because $>$ 10 MeV gammas received at sea level are primarily produced from decays of neutral mesons and electromagnetic showers in the upper atmosphere and compared to neutrons, there is a much larger attenuation of flux of those gammas by the Earth's atmosphere.  
\subsection{Photonuclear cross-sections}
 There is no reported measurement of photonuclear tritium production in Si and Ge, but there is a measurement \cite{shibata1987photonuclear} for photonuclear yield of  \isotope{Co}{60} in copper. Photonuclear yield measurements typically use bremsstrahlung photons produced from high-$Z$ target material using a monoenergetic electron beam. \\
 The yields are reported as integrated cross-section (mb) per quanta for a given bremsstrahlung spectrum with end point energy E$_0$ and can be expressed as:\\
\begin{equation}
 Y(E_0) = \frac{\int_0^{E_0} \sigma(k) N(E_0, k) \, dk}
{\frac{1}{E_0} \int_0^{E_0} k N(E_0, k) \, dk}
\label{eq:2}
\end{equation}
where $\sigma(k)$ is the cross-section, $k$ is a function of photon energy and $N(E_0, k)$ is bremsstrahlung energy spectrum. 
Currie et al. \cite{currie1970photonuclear} reported measurements of photo-tritium yields using a bremsstrahlung spectrum produced by a 90 MeV electron beam in several materials, including Al ($Z=13$) and Zn ($Z=30$). Since $Z$ and $A$ of Al and Si ($Z$=14) and Zn and Ge ($Z$=32) are similar, we consider the measured yields for Al and Zn to be that of Si and Ge, respectively. Assuming 1/$E$ shape of the bremsstrahlung spectrum (up to the end point of 90 MeV) and using  cross-sections, we can also obtain tritium yields using Equation \ref{eq:2} and compare with measurement. However, TALYS doesn't reproduce the measurement well. From measurement, we expect the photo-tritium yield to be about a factor of 10 smaller in Ge compared to Si, however, TALYS underestimates the yield significantly in Silicon. So, to obtain the photo-tritium production in Ge and Si, we assume same shape of cross-section vs energy curve as given by TALYS but scale the curve to match TALYS-given yield to measured yields (The approach taken by  Saldanha et al. \cite{saldanha2020cosmogenic}) . A similar approach, but using the scale factor based on measured yield for end-point energy 130 MeV, is used for \isotope{Co}{60} yield in copper. The comparison of yields between models and measurements are shown in Table \ref{table : photocomparison} in   Appendix \ref{photonuclear_yield_subsection}. Saldanha et al.'s calculation, using scaled TALYS cross-sections and EXPACS-given sea-level gamma flux for NYC, 2003., gives 0.73 atoms/kg/day) using scaled TALYS cross-sections and EXPACS-given sea-level gamma flux for NYC, 2003. Despite the similar approach, we obtain 0.67 atoms/kg/day, the difference is due to the differences in cross-sections at energies $>$ 200 MeV in the newer TALYS compared to the one used by Saldanha et al.  

 \begin{figure}[ht!]
% \centering
 
\subfloat[]{
	\label{subfigure:a}
	\includegraphics[width=0.45\textwidth]{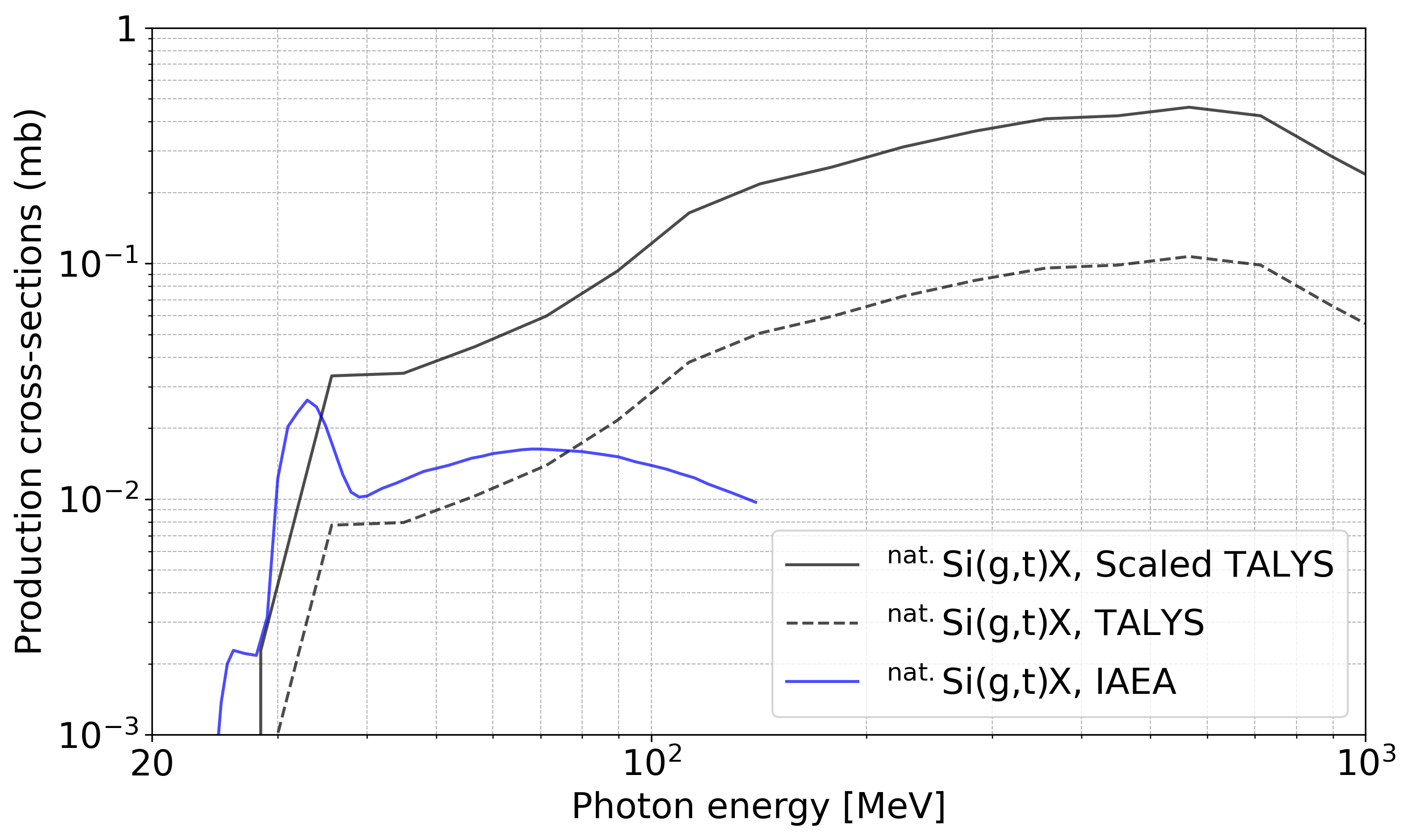}} 
\hspace{0.001\textwidth} % Optional horizontal spacing
\subfloat[]{
	\label{subfigure:b}
	\includegraphics[width=0.45\textwidth]{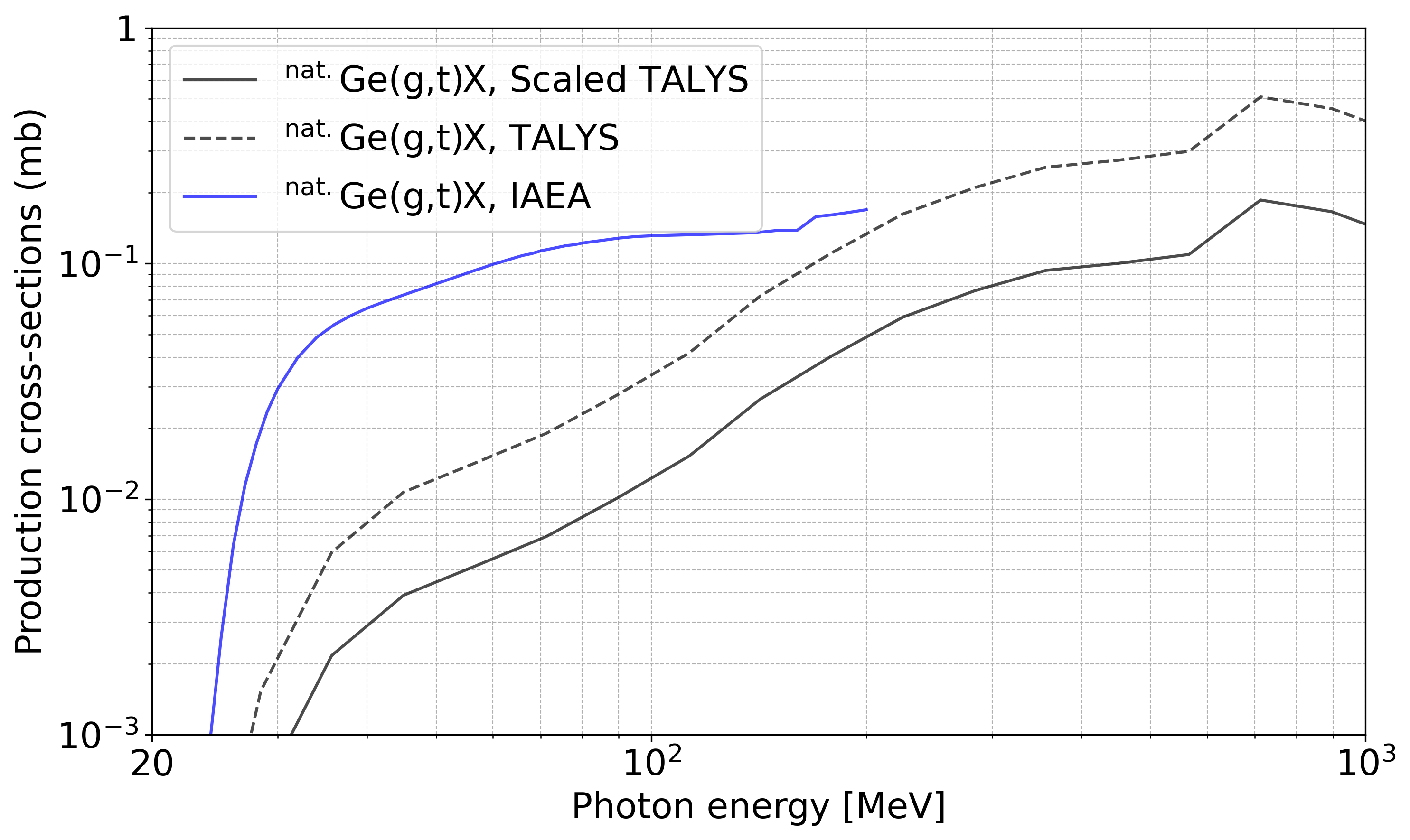}} 
\hspace{0.01\textwidth} % Optional horizontal spacing
\subfloat[]{
	\label{subfigure:c}
	\includegraphics[width=0.45\textwidth]{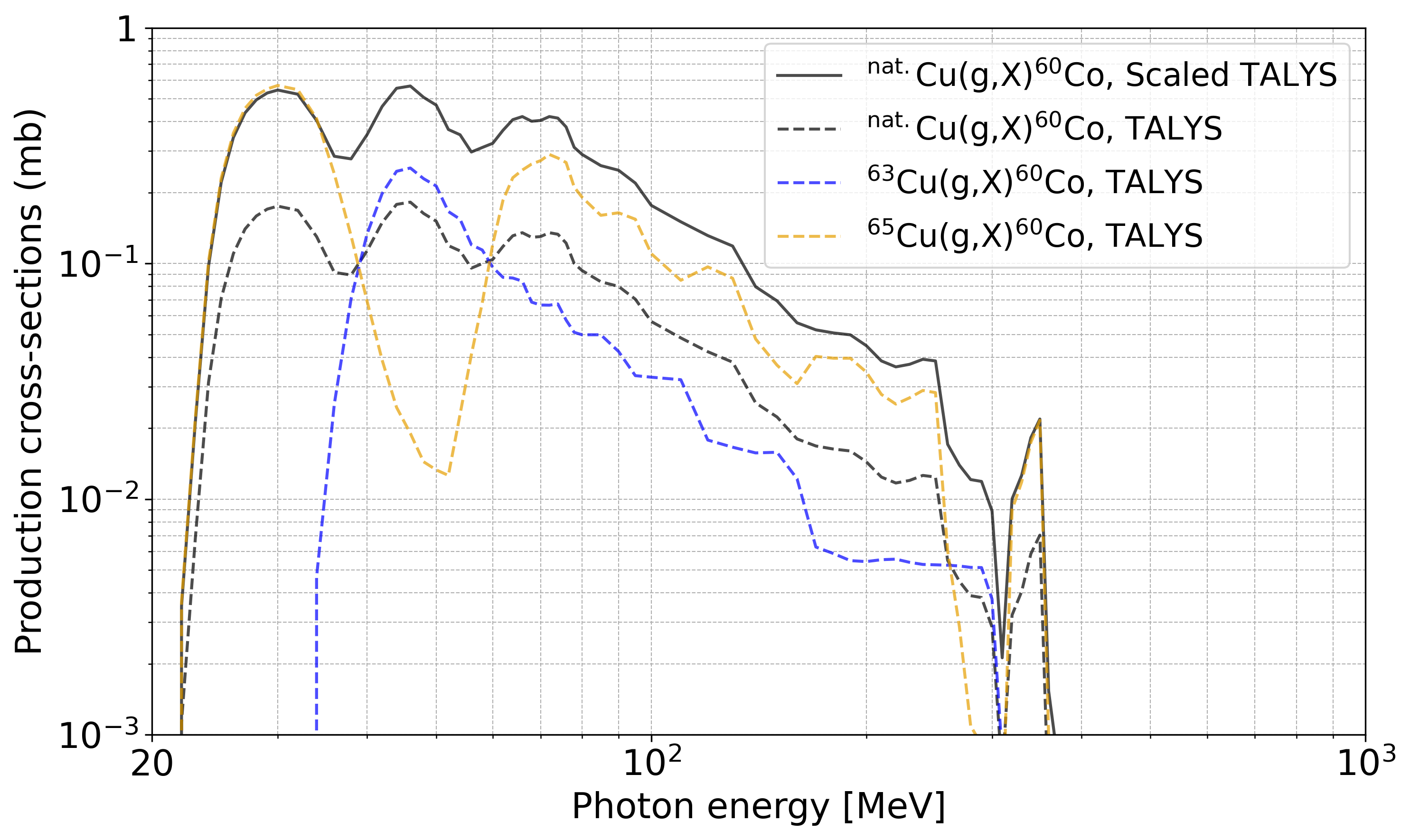}} 
 
\caption{Cross-sections for photon-induced tritium production in (a) Si, (b) Ge and (c) \isotope{Co}{60} production in copper. For tritium production in Si (and Ge), the cross-sections from TALYS, IAEA are also shown for comparison. The TALYS-given cross-section curves are scaled to match the yield measurements and used it to calculate photon-induced activation rates.} 
 \label{fig:photo-cross-sections} 
 \end{figure}

\begin{table}[ht!]
\small
\begin{threeparttable}
\begin{tabular}{|p{3.0cm}|p{1.5cm}|p{1.5cm}|p{1.5cm}|}
\hline
\multirow{2}{*}{Location} & Si & Ge & Cu \\ \cline{2-4}
 & Tritium & Tritium & \isotope{Co}{60} \\ 
\hline
Sea-level        & 0.73 & 0.044 & 1.5 \\
\hline
SUF Tunnel A/C (15--20 mwe)   & $4.4 \times 10^{-2}$ & $2.7 \times 10^{-3}$ & $8.6 \times 10^{-2}$ \\
\hline
PNNL (SUL 30 mwe)        & $2.8 \times 10^{-2}$ & $1.8 \times 10^{-3}$ & $5.0 \times 10^{-2}$ \\
\hline
SLC Adit Storage (50--60 mwe) & $1.5 \times 10^{-2}$ & $9.8 \times 10^{-4}$ & $2.8 \times 10^{-2}$ \\
\hline
\end{tabular}
\end{threeparttable}
\caption{Gamma-induced activation rates (atoms per kg per day.}
\label{table:gamma_production_rates}
\end{table}

Based on these results, phototritium production contributes about 10$\%$ (2$\%$) to the total tritium production in Si (Ge) at shallow depths. In comparison, photo-\isotope{Co}{60} production contributes about 20$\%$ to the total \isotope{Co}{60} production. The contribution is comparable to that from neutrons, which isn't surprising given that the $>$ 10 MeV photon flux is at least two orders of magnitude higher than that of neutrons and a broader resonance in the photon-induced cross-sections for \isotope{Co}{60} production is observed in the energy regime where the photon flux is most significant. We expect the uncertainty in the photon-induced production rates for \isotope{Co}{60} is about 50~\% (scaling TALYS cross-sections based on the measured yield for other end-point energies would change the production rate by only 20\%) and  about 100$\%$ tritium production in for Ge and Si, dominated by the systematic uncertainty in the photonuclear cross-sections.

\section{Other processes}
\label{Other processes}

In this section, we briefly summarize the estimated production of tritium in Ge and Si and \isotope{Co}{60} production in copper from processes other than negative muon capture, neutron and photon interaction.
Cosmic-ray muons and secondary particles, gammas and neutrons, can produce protons and heavier charged particles (deuteron, alpha, \isotope{He}{3}), that can produce tritium in Ge and Si and \isotope{Co}{60} in copper. The tritium production cross-sections for these particles are similar (within a factor of 3) but their flux is smaller \cite{poudel2024subsurface}. In addition, the flux of heavier ions is an order of magnitude or more smaller than the proton flux \cite{poudel2024subsurface}. So, the contribution from other heavy ions is negligible compared to protons. The simulated fluxes of other particles (proton, $\pi^{-}$, $\pi^{+}$) in Ge crystal at 20 mwe depth are discussed in Appendix \ref{Particle_fluence}. The flux of $>$ 10 MeV protons is a factor of 5 smaller than that of $>$ 10 MeV neutrons.  Using the proton flux spectrum  and the TALYS-given cross-sections (shown in Appendix \ref{Particle_fluence}, we obtain the proton-induced tritium production in natural Ge of 2 $\times$ 10$^{-2}$ atom per kg per day, which is about 25$\%$ of neutron-induced tritium production. This level of contribution is expected to be conservatively high, since, as in the case of neutrons, TALYS overestimates the tritium production cross-sections for protons at high energies. For comparison, at 450 MeV, the TALYS-predicted cross-section is a factor of 4 higher than the measured value. However, at 160 MeV, the proton-induced tritium production is about 30 mb, which is a good agreement to measurements \cite{konobeyev1993tritium} (At that energy, for all nuclei from A=30 to 150, the proton-induced tritium production cross-sections are 10-30 mb). 

Neutron production by direct muon spallation is negligible compared to other processes at all depths \cite{wang2001predicting}; therefore, we expect it to be negligible in tritium production in Ge and Si and \isotope{Co}{60} production in copper. Measurements of isotope production from direct muon spallation show the total isotope production rate is dominated by other secondary processes resulting from the direct muon spallation and not the spallation itself \cite{sakurai2024production}.

The cosmic-ray muon–induced interactions in material can also produce pions. Neutral pions, $\pi^{0}$, do not produce isotopes directly, as they decay into energetic gamma rays. Charged pions can produce tritium through $\pi^{-}$ capture or spallation and $\pi^{+}$ spallation; however, these processes are highly suppressed compared to neutron-induced production \cite{mechtersheimer1978measurement,shortt1981cavity}, for instance, by a factor of 10 (20) in C (O) for $\pi^{-}$ capture. Our calculation using the cross-sections calculated using INCL [67-69]\cite{cugnon1982intranuclear,cugnon1997improved,boudard2002intranuclear} (shown in Appendix \ref{Particle_fluence} for tritium production from $\pi^{-}$ and $\pi^{+}$ in natural Ge and using the fluence spectrum for the pions shown in Figure \ref{Figure: proton_and_pionflux in_Ge} gives the tritium production rate (atoms per kg per day) from $\pi^{-}$ and $\pi^{+}$ to be 2.4 $\times$ 10$^{-2}$ and 1.8 $\times$ 10$^{-2}$, which also suggest the contribution to the tritium production from protons and pions is similar. This is reasonable given that the fluence is similar while the cross-sections for charged particle emission \cite{lindgren1978systematics} are similar of pions to that of photons and protons in a broad delta resonance energy regime (typically above 60 MeV)  for pion interaction with nuclei. 

Based on these numbers, we conservatively assume, production from protons, heavy ions and pion capture/spallation to contribute about 80$\%$ relative to the neutron-induced production. At the depth of SUF Tunnel C (20 mwe), this is about 30$\%$ percent of the total tritium production in  Ge while it is 20$\%$ in Si. Assuming these secondary processes producing 80$\%$ to the neutron-induced production of \isotope{Co}{60} in copper, the contribution to the total is 10$\%$ percent. We add this 80$\%$  contribution to that of neutrons for each depth to get the total production rates in the final table. Given the lack of measurements for comparison, we expect the production rates from these secondary processes to be accurate within a factor of 2. Based on this study, the contribution from these processes contribute 10-30$\%$ to the total production rates.

The \isotope{Co}{60} production through \isotope{Cu}{63}(n,$\alpha$)\isotope{Co}{60} has small contribution from $<$ 10 MeV neutrons. Using the SNOLAB neutron flux spectrum \cite{agnese2017projected,villano2024liquid} (normalized with the fast neutron ($>$ 10 keV) flux of $\sim$ 4000 neutrons/cm$^2$/sec (taken from \cite{snolab_handbook}) and the TALYS-given cross-sections, we obtain radiogenic \isotope{Co}{60} production of 6 $\times$ 10$^{-4}$ atoms per ton of copper per year. The radiogenic neutron flux can vary by a factor of 2-4 depending on the composition and U/Th content of the rock, but the production is still negligible - at least five orders of magnitude smaller than the total production rate reported in Section \ref{Result_and_Discuss}, which suggests radiogenic production of  \isotope{Co}{60} is subdominant to cosmogenic to significant depths. 

\section{Results and Discussion}
\label{Result_and_Discuss}
 This paper presents our calculation of the rates of tritium production in Ge and Si, as well as \isotope{Co}{60} production in copper, at shallow depths considering various processes including neutron interactions and negative muon capture. The total production rates for the shallow-depth sites SUF Tunnel A/C, PNNL SUL, and SLC Adit Storage are shown in Table \ref{table:total production rate}.
 
 \begin{table}[H]
\tiny
\footnotesize  % Reduce text size
\setlength{\tabcolsep}{3pt}  % Reduce column padding
\renewcommand{\arraystretch}{0.8}  % Reduce row height
\centering
\scalebox{0.9}{
\begin{tabular}{|p{1.5cm}|p{2.0cm}|p{2.5cm}|p{2.0cm}|}
 \hline
 Location  & Tritium production in Si & Tritium production in Ge & \isotope{Co}{60} production in Cu \\
 \hline
 Sea-level &  125 (TALYS+INCL) \newline 112$\pm$24 \newline(R.Saldanha et.al 2020) & 94 (TALYS+INCL) \newline 74$\pm$9(CDMSLite 2019),  \newline 82$\pm$21 \newline(EDELWEISS-III 2017) & 46 \newline (TALYS, S. Cebrian), \newline 
$39.7 \pm 5.7$ \newline (She et al. 2021), \newline $29.4^{+7.1}_{-5.9}$ \newline(Baudis et al. 2015), \newline $86.2 \pm 7.6$ \newline(Laubenstein et al. 2009). 
 \\ 

\hline
 SUF Tunnel A/C (15-20 mwe) & 0.48 &0.24  & 0.53
 \\
 \hline
 PNNL SUL (30 mwe) & 0.27   & 0.14 & 0.28 \\
 \hline
 SLC Adit Storage (50-60 mwe) & 0.12  & 0.065 & 0.13 \\
 \hline
\end{tabular} }
\caption{Total production rate (atoms per kg per day)} 
\label{table:total production rate}
\end{table}

We find that the tritium production rates in Si (Ge) at SUF Tunnel C is 0.48 (0.24) atoms per kg Ge (Si) per day. This results in the suppression factor of about 250 for Si and 400 for Ge for 20 mwe depth.

 The production suppression factors and a power-law fit on the suppression factors vs depth are shown in Appendix \ref{Production suppression factor vs depth}).  The suppression factors are accurate within a factor of about 2 (See Appendix \ref{Systematics uncertanity}). The relative contribution of various processes to the total production and systematic uncertainties are discussed in Appendix \ref{Systematics uncertanity}. 

 This study shows that tritium production in Ge from negative muon capture and neutron-induced production are comparable even at the shallow depth of about 20 mwe (approx. depth of SUF Tunnels). However, the neutron-induced tritium production in Si only becomes comparable to negative muon-capture at much deeper depth 50 mwe (approx. depth of SLC Adit Storage). While negative muon-capture dominates the \isotope{Co}{60} production in copper at shallow depths at depths at least up to 80 mwe. 
 
 In this paper we have presented a methodology for estimating the production of tritium and \isotope{Co}{60} in Ge/Si and Cu, respectively.  To demonstrate the utility of the calculation methodology, in Table \ref{tab:activity} we present a representative estimation of the activation seen by SuperCDMS components, under plausible exposure scenarios.  Accounting for both production and decays during $\sim$6 years of storage at the tunnel, the induced tritium specific activity is about 1.6 $\mu$Bq/kg in Si and 0.80 $\mu$Bq/kg in Ge (in comparison, unmitigated activation at sea-level would be about 400 $\mu$Bq/kg for Si and 300 $\mu$Bq/kg for Ge).
Similarly, we present in Table \ref{tab:activity} the estimate of induced \isotope{Co}{60} activity in the electroformed copper that has been underground at the PNNL SUL since $\sim$ 2010 i.e for about 15 years. Those copper pieces, with intrinsically low \isotope{Co}{60} activity, can be used for controlled aboveground exposure to measure the \isotope{Co}{60} sea-level production rate, as well as that of other cosmogenically activated isotopes \cite{she2021study}, some of which have uncertainties of a factor of 2 or more.

\begin{table}[H]
\centering
\begin{tabular}{|p{1.2cm}|p{1.2cm}|p{1.3cm}|p{1.3cm}|p{1.3cm}|}
\hline
Location & Duration \newline (years) & Tritium in Si ($\mu$Bq/kg) & Tritium in Ge ($\mu$Bq/kg) & \isotope{Co}{60} in Cu ($\mu$Bq/kg)\\
\hline
SUF Tunnels A/C & 6 & 1.6 & 0.80 & 3.3 \\
\hline
PNNL SUL & 15 & 1.8 & 0.92 & 2.8\\
\hline
SLC Adit Storage & 1 & 0.076 & 0.041 & 0.19  \\
\hline
\end{tabular}
\caption{Induced activities of Tritium in Si and Ge, and \isotope{Co}{60} in Cu at different shallow-depth sites for considered exposures.}
\label{tab:activity}
\end{table}
\FloatBarrier

Our results show that shallow-depth facilities such as SUF and PNNL can be used for Ge and Si crystal growth, long-term detector storage, and detector fabrication and assembly when detector sensitivity to rare-event searches is limited by elevated cosmogenic activation during aboveground exposure. Future large-scale rare-event search experiments may also use available shallow-depth facilities for the storage of detector and shielding materials, and potentially carry out detector testing and assembly underground.

\section{Acknowledgments}
We thank the SuperCDMS Collaboration for their support and for sharing the results of muon witness detector measurements, as well as the resulting estimate of the mwe depth of the SLC Adit Storage. We thank Dr. Saldanha Saldanha from Pacific Northwest National Laboratory (PNNL) for helpful discussions on various aspects of the physics. We also thank Prof. Shawn Westerdale from the University of California, Riverside for valuable discussions on muon-induced tritium production. We thank Sanjay Sharma Poudel from the University of Houston for discussions related to FLUKA simulations and the INCL nuclear code. We thank Prof. Pekka K. Sinervo from the University of Toronto for carefully reading the manuscript and providing valuable feedback and comments.

The author, Sagar S. Poudel, dedicates this work to his late mother, SatyaDevi Bastola.
\bibliography{apssamp}% Produces the bibliography via BibTeX.

\section{Appendix}\label{Appendix}

\subsection{Sea-level cosmic-ray neutron flux}
\label{sea_level_neutrons}
\begin{figure}[h]
\centering
\includegraphics[width=1.0\linewidth]{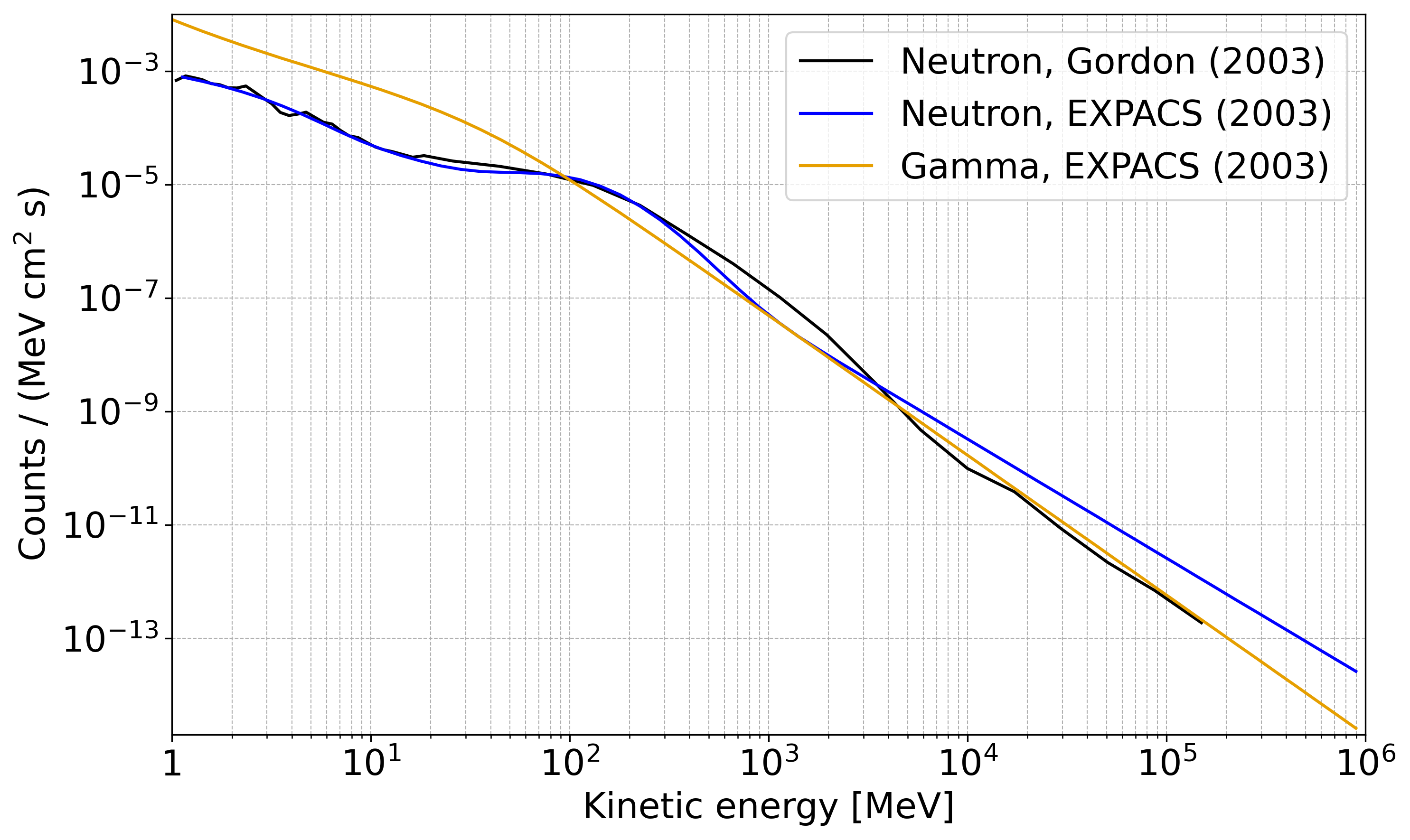}
\caption{Differential neutron flux (Gordon vs EXPACS) obtained for NYC,2003. EXPACS-given gamma flux is also shown for comparison} 
\label{Figure: Neutron_comparison}
\end{figure}

Figure \ref{Figure: Neutron_comparison} shows the differential neutron flux for New York City (2003), with Gordon’s measurements overlaid with that calculated using EXPACS for the same location. The neutron flux above 10 MeV from EXPACS (3.3 $\times$ 10$^{-3}$ cm$^{-2}$s$^{-1}$) is in good agreement with Gordon’s value (3.5 $\times$ 10$^{-3}$ cm$^{-2}$s$^{-1}$). Also gamma flux as obtained using EXPACS for the same location. 
%\subsubsection{Neutron flux}
\subsection{Neutron flux attenuation}
\label{neutron_flux_attenuation}
We generate CRY-code-given neutrons in 100 m $\times$ 100 m plane and propagate through our modeled rock (of density 2.7 g/cm$^3$) and record $>$ 10 MeV neutron flux at various depths. Aboveground neutron spectrum used as input in the simulations was generated  for NYC, 2003. Since the CRY-given $>$ 10 MeV neutron flux (1.61 $\times$ 10$^{-3}$) is smaller than the flux measured by Gordon et al\cite{gordon2004measurement}, the flux at various depths are obtained by normalizing the simulation results to  Gordon's $>$ 10 MeV neutron flux. The results for various depths in the modeled rock are shown in Table \ref{table: aboveground-neutron-flux}. For 4 m (11 mwe) depth, we obtain the $>$ 10 MeV neutron flux of (1.5 $\pm$ 0.4) $\times$ 10$^{-6}$, a factor of 2300 smaller than at sea-level. 
Comparing the neutron flux  with muon-induced neutrons flux in Table \ref{table: aboveground-neutron-flux}, the flux of $>$10 MeV neutron is already dominated by muon-induced interactions for depth 10 mwe or greater. This is also observed in the simulation results on LNGS rock shown in \cite{vanhoefer2015neutron}.   
\begin{table}[ht!]
 \small
\begin{tabular}{|p{2.0cm}|p{2.5cm}|p{3.0cm}|}
 %\multicolumn{5}{}{Major $\beta$-decay channels in Ar42 chain}\\ 
 \hline
 Depth (m)& Depth (mwe) & $>$ 10 MeV atmospheric neutron flux\\
 \hline
 0 (sea-level) & - & 3.50 $\times$ 10$^{-3}$ \\
 \hline
 1  & 2.7 & (5.4 $\pm$ 0.05) $\times$ 10$^{-4}$\\
 \hline
 2  & 5.4  & (7.9 $\pm$ 0.2)$\times$ 10$^{-5}$ \\
 \hline
 3 & 8.1 &  (9.6 $\pm$ 0.9) $\times$ 10$^{-6}$\\
 \hline
 4 & 10.8 & (1.5 $\pm$ 0.4) $\times$ 10$^{-6}$\\
 \hline
 \end{tabular}
  \caption{Flux of $>$ 10 MeV atmospheric neutrons at various depths.}
  \label{table: aboveground-neutron-flux}
  \end{table}

\subsection{Cosmogenic neutron flux (Crust vs Limestone) at shallow depth}
\label{crust_vs_limestone}
\begin{figure}[h]
\centering
\includegraphics[width=1.0\linewidth]{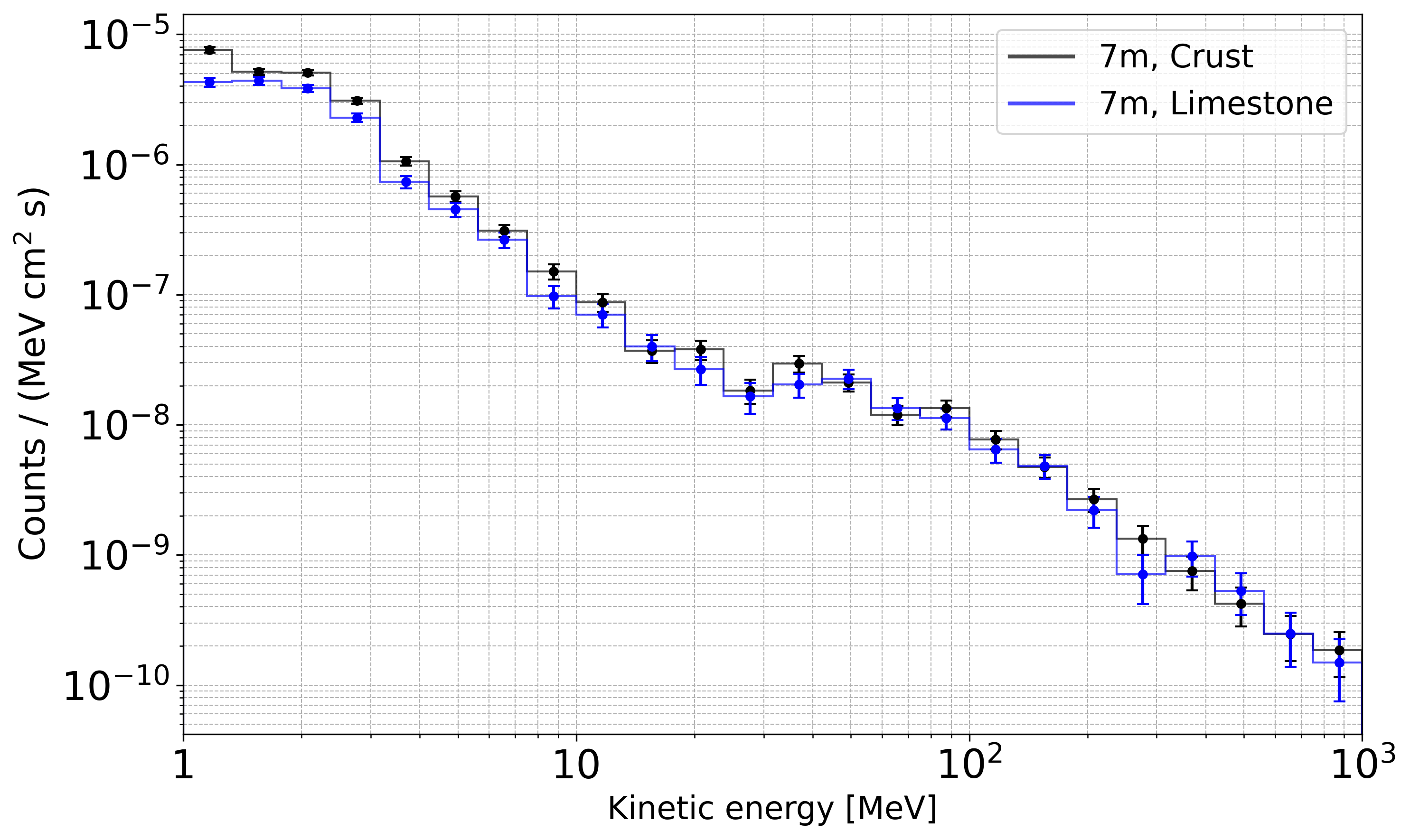}
\caption{Neutron flux spectrum for crust vs limestone composition for 7 m rock overburden} 
\label{Figure: Neutron_comparison_crust_lime}
\end{figure}

We simulate the muon-induced neutron flux assuming a limestone rock overburden while keeping the same density (2.7 g/cm$^3$), and obtain the corresponding neutron flux and energy spectrum. The neutron energy spectrum in limestone at a depth of 7 m, corresponding to the depth of the SUF Tunnels, is shown in Figure \ref{Figure: Neutron_comparison_crust_lime}. The spectrum for the crust composition with a 7 m overburden is also shown for comparison. In addition, the neutron flux in the energy ranges 2–10 MeV and $>$ 10 MeV for the two rock compositions is reported in Table \ref{table: neutron-flux-composition}.
In the energy range 2–10 MeV, where neutrons from muon capture are important, the neutron flux for the limestone composition is about 40\% smaller, while the $>$ 10 MeV neutron flux is about 10\% smaller. As shown in Figure \ref{Figure: Neutron_comparison_crust_lime}, the spectra become increasingly similar at higher neutron energies, where hadronic showers are the dominant contributor \cite{malgin2017energy}.
\begin{table}[ht!]
 \small
\begin{tabular}{|p{2.0cm}|p{3.0cm}|p{3.0cm}| }
 %\multicolumn{5}{}{Major $\beta$-decay channels in Ar42 chain}\\ 
 \hline
  \textbf{Rock composition} & \textbf{(2-10) MeV neutron flux (/cm$^{2}$/s)} & \textbf{$>$ 10 MeV neutron flux (/cm$^2$/s)} \\
 \hline
 Crust & 7.4 $\times$ 10$^{-6}$ & 3.0 $\times$ 10$^{-6}$\\
 \hline
 Limestone & 5.4 $\times$ 10$^{-6}$ & 2.7 $\times$ 10$^{-6}$\\
 \hline
 \end{tabular}
\caption{Comparison of (2-10) MeV and $>$ 10 MeV cosmic-ray muon-induced neutron flux at the depth of 7 m (about 20 mwe, a depth of SUF Tunnels) for overburdens of two different compositions. }
\label{table: neutron-flux-composition}
\end{table}
\subsection{FLUKA simulations and particle transport}
\label{FLUKA simulations and particle transport}
We use the FLUKA particle transport code \cite{FLUKA_1_battistoni2015overview,FLUKA_2_bohlen2014fluka,FLUKA_3_ballarini2007physics}, which is known for its robust hadronic models required to simulate muon-induced neutron production as well as the production and propagation of secondary particles from muon interactions. We use the FLUKA simulation package (released in 2021) distributed by INFN.

The FLUKA physics settings used are similar to those described in \cite{empl2014fluka,poudel2024subsurface}. We use the PRECISIOn settings as the default in this work for the generation and transport of particles from muon-induced interactions. Photonuclear interactions and bremsstrahlung processes for muons are also enabled. We use the COALESCE and EVAPORAT options to enable heavy-ion production and nuclear evaporation in the simulations. Full ion transport is enabled using the IONTRANS option. For neutrons with energies below 20 MeV, the default cross-sections from FLUKA’s dedicated low-energy neutron libraries are used, which rely on evaluated nuclear data files or experimental measurements where available.

\subsection{$>$ 10 MeV neutrons in SUF Tunnel C and PNNL SUL}
\label{back-injected neutrons}
In this work, we built a model for SUF Tunnel C and PNNL SUL and ran separate simulations to investigate the $>$ 10 MeV neutron flux at these sites. In the simulations, SUF Lab is implemented as a cavern with dimensions 3 m (l) × 15 m (b) × 3 m (h) under 7 m of rock overburden, while PNNL SUL is modeled as a cavern with dimensions 10 m (l) × 10 m (b) × 4 m (h) under 12 m of rock overburden. These dimensions are representative of the actual cavern geometries at these underground sites. In both cases, the rock overburden is assumed to be flat. The rock density and composition of the overburden are taken to be the same as those of crustal rock. In the simulations, we record neutrons entering the cavern through the ceiling, walls, and floor.
The mean energy of $>$ 10 MeV neutrons and the relative percentages of neutrons entering the caverns through the ceiling, walls, and floor are shown in Table \ref{tab:back_scatter_neutrons}. 
\begin{table}[]
\scriptsize
\begin{tabular}{|c|c|c|c|}
\hline
\multicolumn{1}{|l|}{\textbf{Site}} & \multicolumn{1}{l|}{\textbf{Location}} & \multicolumn{1}{l|}{\textbf{\begin{tabular}[c]{@{}l@{}}Mean K.E. (MeV) \\ of \textgreater 10 MeV neutrons\end{tabular}}} & \multicolumn{1}{l|}{\textbf{Ratio}} \\ \hline
Sea level & NYC, 2003 & 170 & -- \\ \hline
\multirow{3}{*}{SUF Tunnel C} & Ceiling & 122 & 0.35 $\pm$ 0.04 \\ \cline{2-4} 
 & Floor & 64 & 0.20 $\pm$ 0.02 \\ \cline{2-4} 
 & Wall & 86 * & 0.45 $\pm$ 0.04 \\ \hline
\multirow{3}{*}{PNNL SUL} & Ceiling & 128 & 0.45 $\pm$ 0.04 \\ \cline{2-4} 
 & Floor & 57 & 0.18 $\pm$ 0.02 \\ \cline{2-4} 
 & Wall & 96 * & 0.37 $\pm$ 0.04 \\ \hline
\end{tabular}
\caption{Mean kinetic energy of $>$ 10 MeV neutrons entering shallow-depth sites of interest, along with the fractions of the total neutron flux entering through the ceiling, floor, and walls. The mean energy of $>$ 10 MeV neutrons at sea level (Gordon’s measurements in NYC) is also shown for comparison. * includes contributions from both through-going and back-injected neutrons.}
 \label{tab:back_scatter_neutrons}
\end{table}

\begin{figure}[h]
\centering
\includegraphics[width=1.0\linewidth]{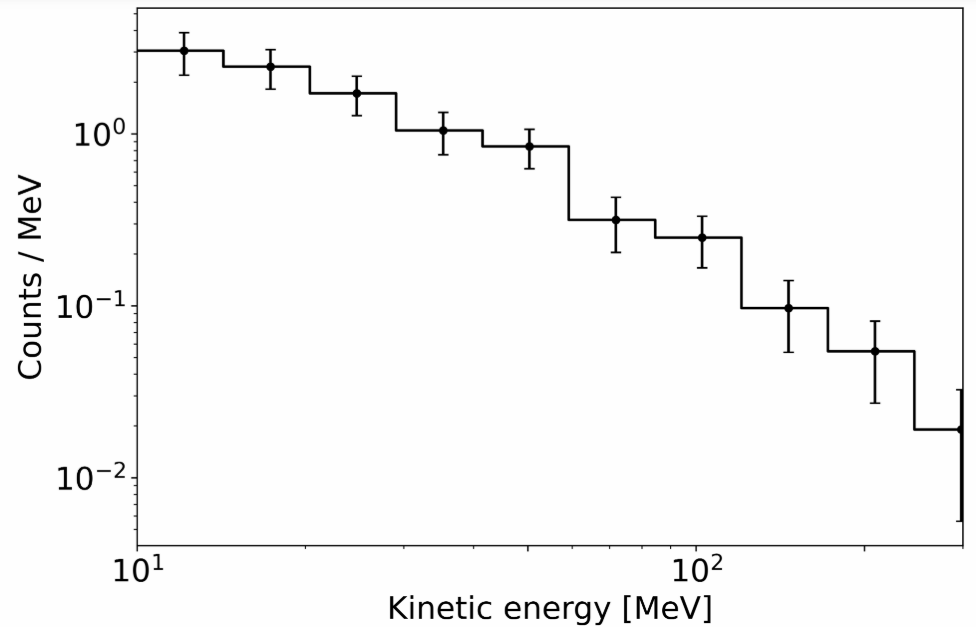}
\caption{The spectrum of $>$ 10 MeV neutrons from the floor. The spectrum is softer compared to that of through-going neutrons with mean neutron energy being 60 MeV.} 
\label{Figure: floor_neutrons}
\end{figure}

The results show that the energy spectrum of $>$ 10 MeV neutrons entering shallow-depth sites is much softer than at sea level. The $>$ 10 MeV neutrons entering from the floor have a significantly lower mean energy. Since both the mean energy and the ratio of floor to ceiling (and total) neutrons are consistent within 1 $\sigma$ binomial uncertainty for the SUF and PNNL sites, the results for both sites are combined and shown in Figure \ref{Figure: floor_neutrons}.

Based on the simulations, distinguishing through-going from back-injected neutrons entering through the walls is more ambiguous. The reported mean energy for wall-entering neutrons includes some contamination from back-injected low-energy neutrons. Otherwise, we expect the energy spectrum of through-going neutrons from the walls to be similar to, or slightly harder than, that from the ceiling, since wall-entering neutrons are produced by muons traveling at larger slant angles. The mean muon energy at SUF is about 10 GeV for muons entering through the ceiling and about 12 GeV for those entering through the walls.

Assuming that the ratio of through-going to back-injected neutrons for the walls is the same as that inferred for the ceiling-to-floor case, we estimate that about 20\% of wall-entering neutrons are back-injected, similar to the fraction from the floor. We therefore add 40\% to the production rates obtained in rock to account for enhanced activation from back-injected neutrons. This value may be conservative, since the neutron spectrum of back-injected neutrons is softer, although it can also depend on cavern geometry. A study of cosmic-ray–induced neutrons without energy restrictions \cite{vanhoefer2015neutron} suggests a back-injected-to-downward neutron flux ratio of approximately 1:1. Based on that study and our results, we expect the uncertainty in back-scatter neutron-induced production to be $\leq$ 30\%.

\subsection{Muon stopping rate at SUF Tunnel C}

\label{stopping_muon_at_SUF}
In this work, we build a model of SUF Tunnel C (15 m (b) $\times$ 3 m (l) $\times$ 3 m (h)) under 7 m of rock overburden with the same composition and density as shown in Table \ref{table: rock_composition}. A 7 m rock overburden corresponds to a depth of 20 mwe. for our modeled rock. Using the CRY code, we generate $1.2 \times 10^{8}$ negative muons on a 100 m $\times$ 100 m plane and use FLUKA simulations to propagate the muons through the overburden and record their interactions.

In the simulations, we place one Ge and one Si cylinder (each of radius 25 cm and thickness 10 cm) in SUF Tunnel C, separated sufficiently to avoid mutual shielding. The cylinder thickness is small enough that the muon energy spectrum can be assumed uniform across their volumes. We record the muons entering and exiting the cylinders, and the difference in counts gives the number of stopping negative muons\footnote{This technique works for shallow depths. At greater depths, muons can also be produced locally \cite{cassiday1973calculation}, for example from decays of pions and kaons produced in hadronic showers.}. This yields 62 and 24 stopping muons in the Ge ($\rho = 5.323$ g/cm$^{3}$) and Si ($\rho = 2.33$ g/cm$^{3}$) cylinders, respectively.

Using the number of stopped negative muons in the Ge cylinder, a positive-to-negative muon flux ratio of 1.1, and the CRY-predicted aboveground muon flux of $1.15 \times 10^{-2}$ cm$^{-2}$ s$^{-1}$, we obtain a negative muon stopping rate of $90 \pm 11$ kg$^{-1}$ day$^{-1}$ at SUF Tunnel C. This value is consistent within 30\% with the result obtained in Section \ref{Stopping_muons_and_cosmogenic_activation} using S. Charalambus’s parameterization for 20 mwe, the approximate depth of SUF Tunnel C.

\subsection{FLUKA simulations: Particle yield per stopped negative muon}
\label{Particle_yield_from_stopped_muons_FLUKA}
We also use FLUKA simulations to obtain the tritium yield in Ge and Si and the \isotope{Co}{60} yield in Cu from negative muons. In the simulations, we generate negative muons with energies of 10 keV (for Ge and Cu) and 5 keV (for Si), slightly above the binding energy of a muon in an atomic orbital, uniformly and isotropically within modeled spheres of Ge, Si, and Cu. The spheres, each with a radius of 25 cm, are large enough that losses from escaping muons at the surface are negligible.

We record the produced isotopes using FLUKA’s RESNUCLEi card without transporting secondaries, in order to obtain isotope production from stopping muons in the materials of interest. The RESNUCLEi card records isotopes in a well-defined volume and reports the yield as the number of isotopes produced per cm$^{3}$ per primary particle; in this case, each primary can be treated as a stopped muon, since the initial muon energy is low enough that all muons are assumed to stop within the sphere. For a uniform and isotropic distribution of stopped muons, the isotope yield per negative muon can be obtained by multiplying by the volume of the sphere in which the muons are generated. This approach is equivalent to simulating a point-like isotropic low-energy negative-muon beam in sufficiently thick material and expressing the result as a yield per stopped negative muon.

The results for tritium yield per stopped muon in Ge and Si, \isotope{Co}{60} in Cu , and yield for other activated isotope of general interest are also reported in Table \ref{tab:Yield per stopped negative muon}.

\begin{table}[h]
\tiny
\centering
\begin{tabular}{|c|c|c|c|}
\hline
Material & Isotopes & Yield per stopped negative muon  & Uncertainty (\%) \\ \hline

\multirow{3}{*}{Si} 
 & \textbf{Tritium} & \textbf{3.35 $\times$10$^{-3}$} & \textbf{0.3}  \\ \cline{2-4}
 & \isotope{Be}{7}  & 4.05 $\times$ 10$^{-5}$ & 2 \\ \cline{2-4}
 & \isotope{Na}{22} & 2.98 $\times$10$^{-3}$ & 0.2 \\ \hline

\multirow{7}{*}{Ge} 
 & \textbf{Tritium} & \textbf{1.26 $\times$10$^{-3}$} & \textbf{0.5} \\ \cline{2-4} 
 & \isotope{Co}{60} & 6.42 $\times$ 10$^{-5}$  & 5\\ \cline{2-4}
 & \isotope{Co}{58} & 3.88 $\times$ 10$^{-5}$  & 2 \\ \cline{2-4}
 & \isotope{Co}{57} & 4.8 $\times$ 10$^{-5}$ & 15 \\ \cline{2-4}
 & \isotope{Mn}{54} & 1.0 $\times$ 10$^{-6}$ & 23 \\ \cline{2-4}
 & \isotope{Zn}{65} & 2.69 $\times$ 10$^{-3}$ & 0.3 \\ \cline{2-4}
 & \isotope{Ge}{68} & 5.1 $\times$ 10$^{-6}$ & 12 \\ \hline
 
 \multirow{7}{*}{Cu} 
 & \textbf{\isotope{Co}{60}} & \textbf{4.10 $\times$ 10$^{-3}$}  & \textbf{0.6} \\ \cline{2-4}
 & \isotope{Co}{58} & 6.07 $\times$ 10$^{-3} $ & 0.2 \\ \cline{2-4}
 & \isotope{Co}{57} & 2.64 $\times$ 10$^{-3} $ & 0.4 \\ \cline{2-4}
 & \isotope{Co}{56} & 1.64 $\times$ 10$^{-4} $ & 3 \\ \cline{2-4}
 & \isotope{Fe}{59} &1.03 $\times$ 10$^{-3} $ & 1 \\ \cline{2-4}
 &  \isotope{Mn}{54} &7.94 $\times$ 10$^{-4} $ & 1 \\ \cline{2-4}
 & \isotope{Sc}{46} &8 $\times$ 10$^{-7} $ & 36 \\ \hline

\end{tabular}
\caption{Yields per stopped negative muon obtained from FLUKA simulations. The uncertainties are statistical. Rows in bold correspond to the isotopes of interest in this work. Additional activation isotopes with statistical uncertainties below 50\% are included as supplementary material.}
\label{tab:Yield per stopped negative muon}
\end{table}

\begin{figure}[h]
\centering

\includegraphics[width=1.0\linewidth]{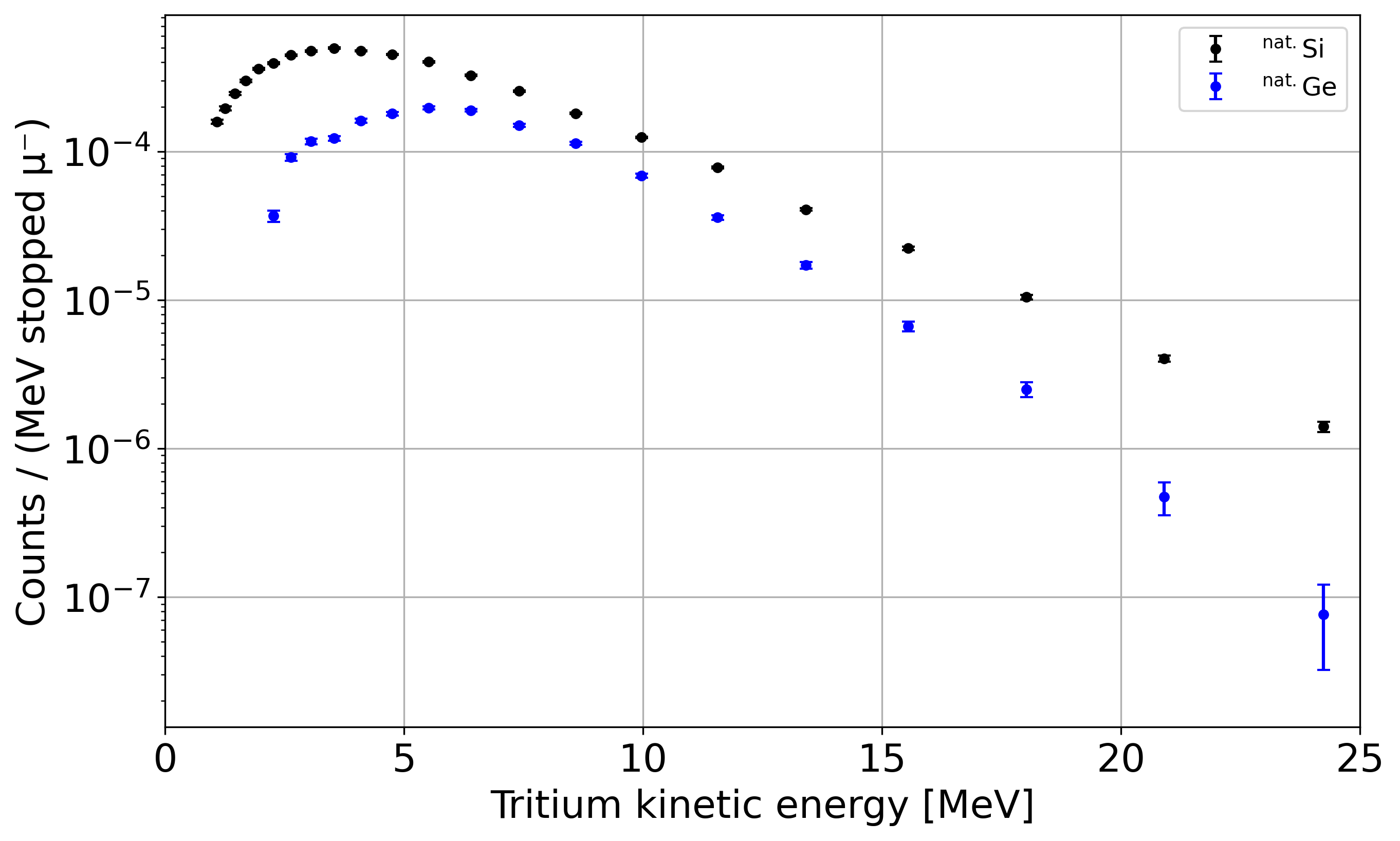}
\caption{Tritium KE spectrum from negative muon capture in natural Ge and Si obtained from FLUKA simulations.} 
\label{Figure: Kinetic_spectrum_tritium}
\end{figure}

We also record interaction-level information using the USRDRAW routine in mgdraw.f to count muon captures and decays (based on interaction codes 101 for capture and 102 for decay) as well as to record the kinetic energy of tritium produced from negative muon capture in Ge and Si. The tritium kinetic energy spectra for natural Ge and Si are shown in Figure \ref{Figure: Kinetic_spectrum_tritium}. Our simulation shows that, although FLUKA uses the Goulard–Primakoff formula \cite{goulard1975relation} for light nuclei and empirical fits for medium and heavy nuclei, it reproduces the measured negative muon capture probabilities in Si, Ge, and Cu (0.66, 0.92, and 0.92, respectively) \cite{suzuki1987total}.

\subsection{Photonuclear yield (Measurements vs Model)}
\label{photonuclear_yield_subsection}
The comparison of photo-tritium yields obtained using TALYS and the IAEA library \cite{kawano2020iaea}, together with measurements, is shown in Table \ref{table : photocomparison}. The measurements of tritium yield are based on a bremsstrahlung photon spectrum with an end-point energy of 90 MeV and are taken from \cite{currie1970photonuclear}. The measurements used for Si and Ge correspond to those for Al and Zn, respectively. Two measurements of the \isotope{Co}{60} yield in Cu, shown in the table, are taken from \cite{shibata1987photonuclear}. Since the IAEA database \cite{kawano2020iaea} does not include cross-sections above the pion-production threshold (140 MeV), the IAEA yields for \isotope{Co}{60} are not available.
\begin{table}[ht!]
%\tiny
\scalebox{0.9} {
\begin{tabular}{|p{2.0cm}|p{2.0cm}|p{2.0cm}|p{2.0cm}|}
\hline
Reaction & Measured yield & Calculated yield (TALYS) & Calculated yield (IAEA) \\ \hline
nat. Si(g,t)
X & 7.2 $\times$ 10$^{-2}$  \newline (for nat. Al) & 1.7 $\times$ 10$^{-2}$  & 2.0 $\times$ 10$^{-2}$  \\ \hline
nat. Ge(g,t)X & 7.0 $\times$ 10$^{-3}$ \newline (for nat. Zn) &  1.9 $\times$ 10$^{-2}$  & 1.1 $\times$ 10$^{-1}$ \\ 
\hline
nat. Cu(g,X)\isotope{Co}{60} & 4.70 $\times$ 10$^{-1}$ \newline 6.02 $\times$  10$^{-1}$ & 1.7 $\times$ 10$^{-1}$ \newline 1.9 $\times$ 10$^{-1}$ & - \\ 
 \hline
\end{tabular} }
\caption{Comparison of photo-tritium yield obtained from TALYS and the IAEA database with measurements. The measurements for Al (used for Si) and Zn (used for Ge) at a bremsstrahlung energy of 90 MeV are taken from \cite{currie1970photonuclear}, while the measurements for Cu at bremsstrahlung energies of 190 and 220 MeV are taken from \cite{shibata1987photonuclear}.}
\label{table : photocomparison}
\end{table}

\subsection{Comparison of particle flux and Cross-sections}
\label{Particle_fluence}
We obtain the fluence of various cosmic-ray–induced secondary particles in a Ge crystal at a depth of 20 m.w.e. The input muon spectrum and simulation strategy are the same as those described in Section \ref{sec:neutron_flux}. In the simulation, we place nine Ge crystals with a radius of 325 cm and a thickness of 10 cm on the floor of an enlarged SUF Tunnel geometry (20 m × 20 m × 3 m) and combine the statistics from all simulation runs. The crystal thickness is chosen to be small and representative of typical experimental detectors. It is not sufficient to allow hadronic shower development within the Ge material; otherwise, the particle flux would be artificially enhanced.

\begin{figure}[h]
\centering
\includegraphics[width=1.0\linewidth]{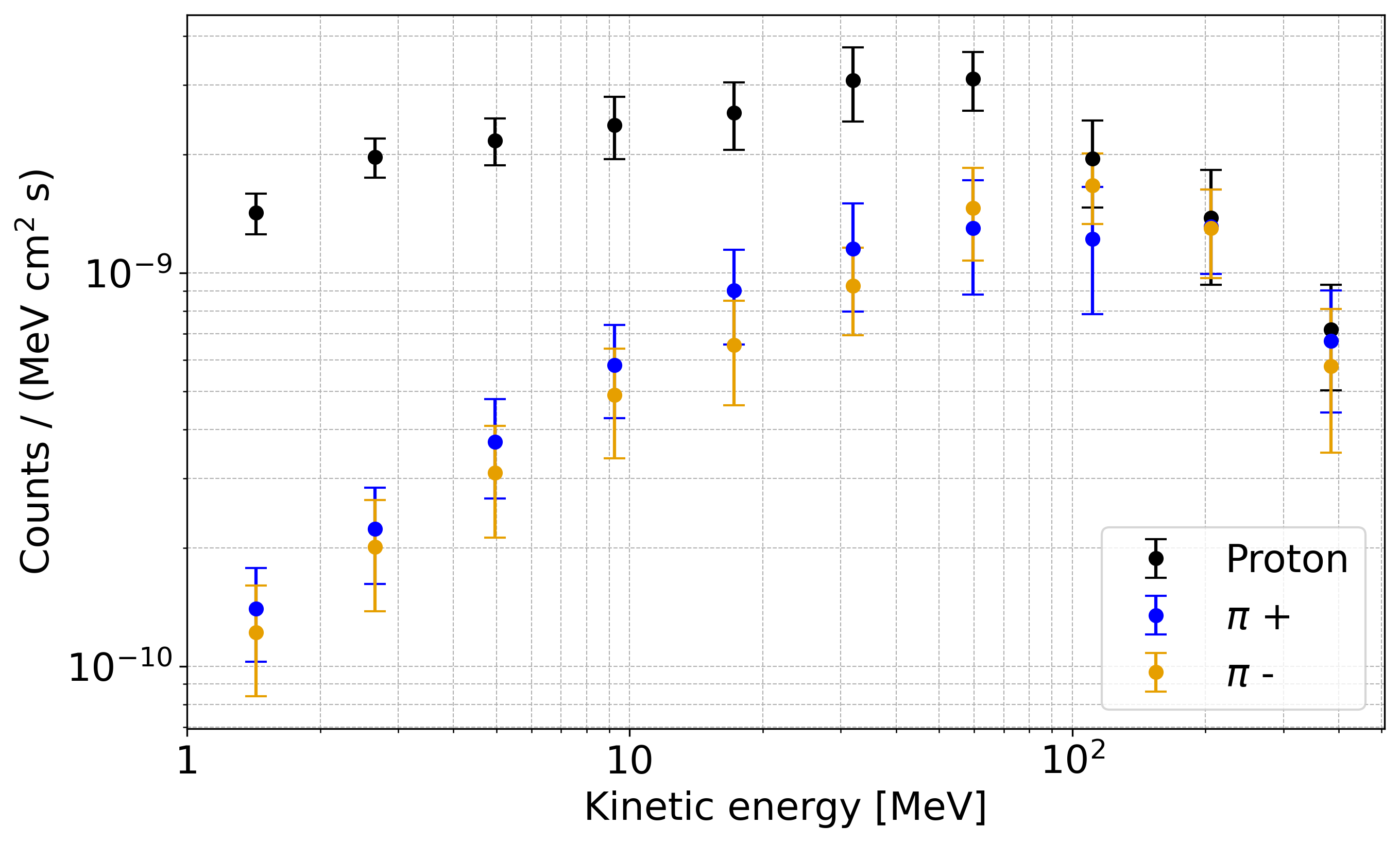}
\caption{Proton and charged pion fluxes in Ge (10 cm thick) at 20 mwe} 
\label{Figure: proton_and_pionflux in_Ge}
\end{figure}

The fluences of protons, $\pi^{+}$, and $\pi^{-}$ are shown in Table \ref{table:fluence rate}, and the corresponding flux spectra are given in Figure \ref{Figure: proton_and_pionflux in_Ge}. The fluence is recorded using FLUKA’s USRTRACK card, which tracks path length and provides a volume-weighted fluence. The normalization is based on the CRY-predicted muon flux at sea level.

\begin{figure}[h]
\centering
\includegraphics[width=1.0\linewidth]{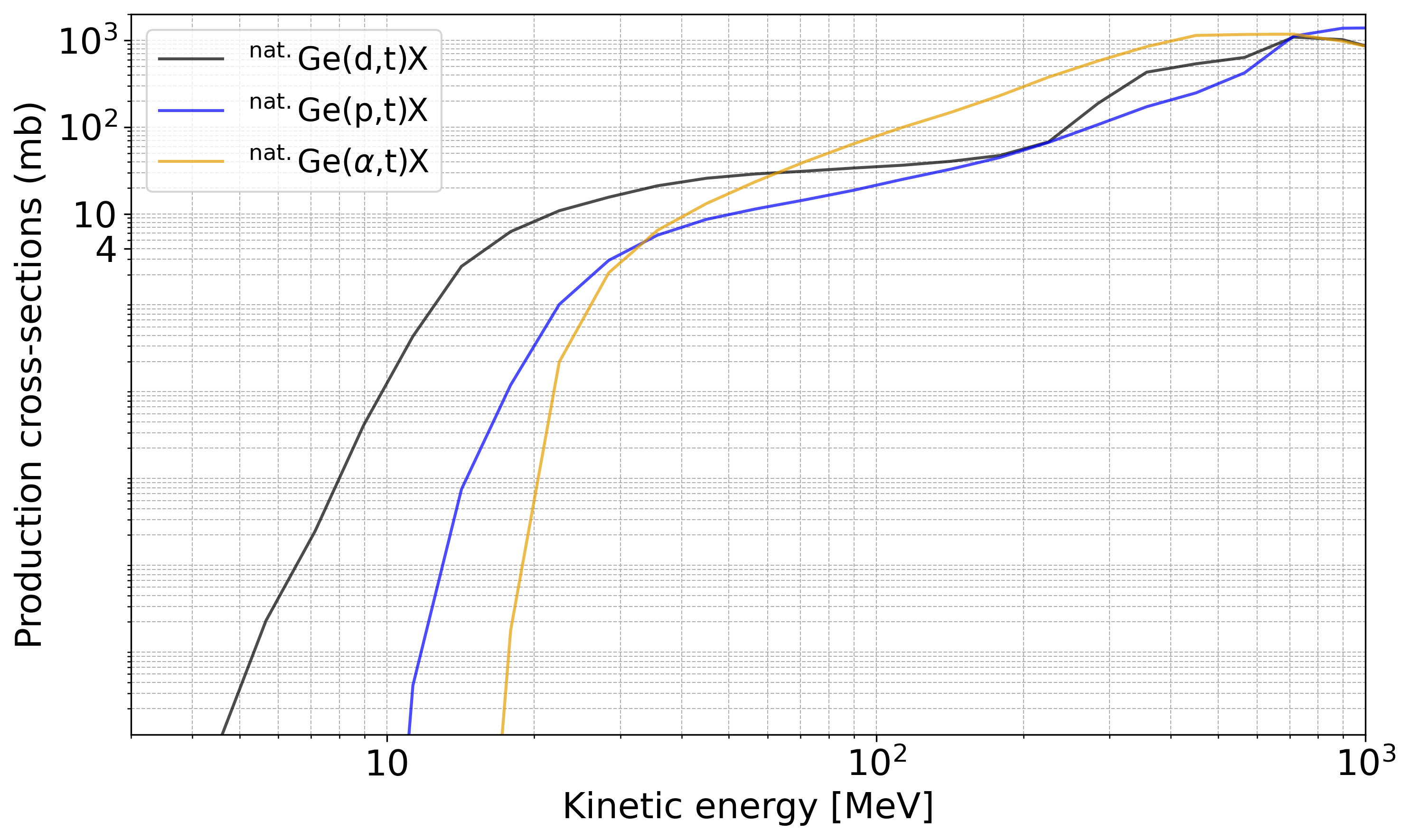}
\caption{Tritium production cross-sections for proton, deuteron and alpha in Ge (obtained using TALYS)} 
\label{Figure: particle_cross-sections_tritium_in_Ge}
\end{figure}

\begin{figure}[h]
\centering
\includegraphics[width=1.0\linewidth]{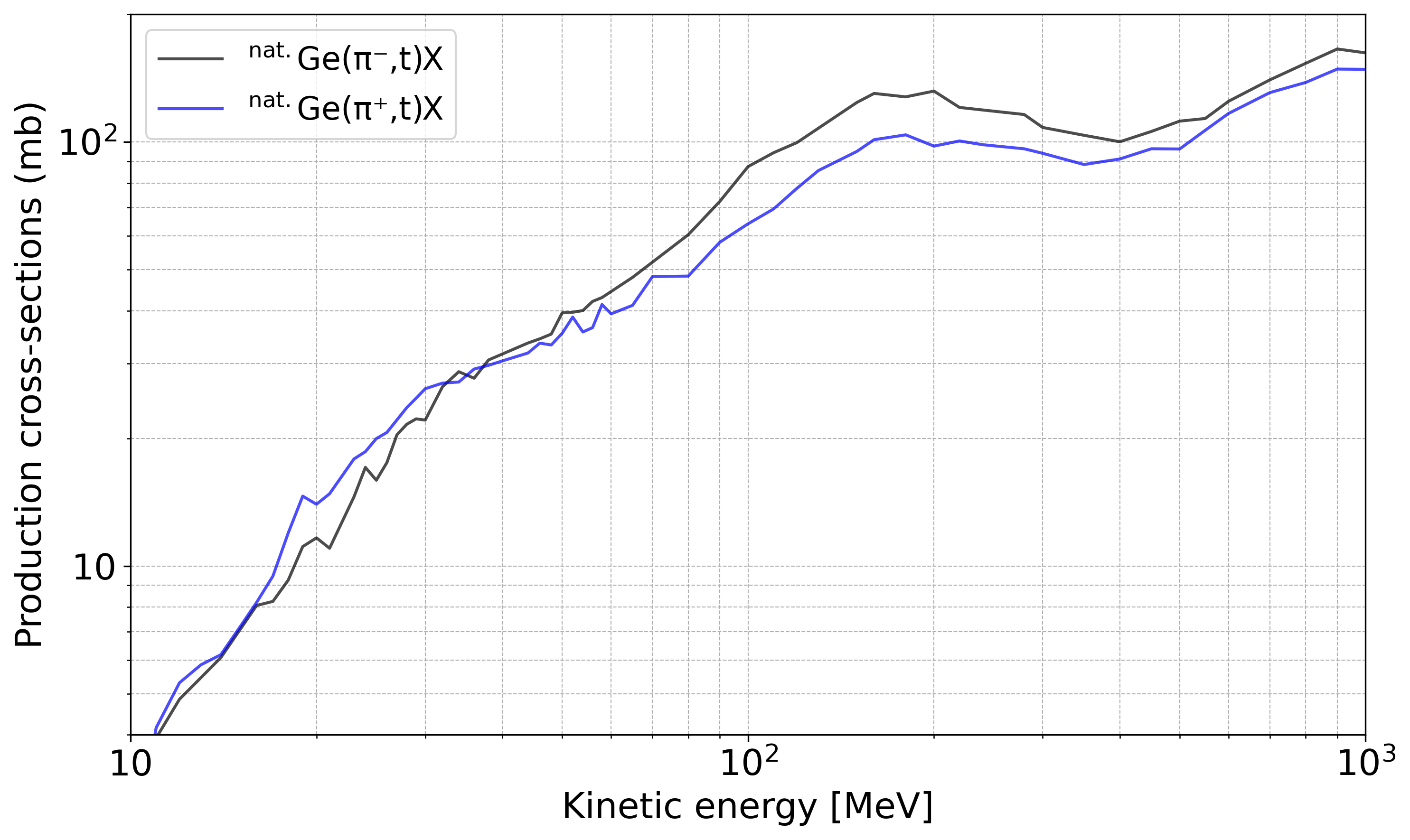}
\caption{Tritium production cross-sections for pions in natural Ge (obtained using INCL)} 
\label{Figure: pion_cross-sections-tritium_in_Ge}
\end{figure}

The tritium production cross-sections in Ge for deuterons, protons, and alpha particles (obtained using TALYS) are shown in Figure \ref{Figure: particle_cross-sections_tritium_in_Ge}, while those for pions (obtained using the INCL model) are shown in Figure \ref{Figure: pion_cross-sections-tritium_in_Ge}.
\begin{table}[ht!]
\footnotesize  % Reduce text size
\setlength{\tabcolsep}{3pt}  % Reduce column padding
\renewcommand{\arraystretch}{0.8}  % Reduce row height
\centering
\begin{tabular}{|p{2.5cm}|p{3.5cm}|}
 \hline
 Particle type  & Flux (/cm$^2$/s) \\
 \hline
 Proton &  (6.9 $\pm$  1.3)$\times$10$^{-7}$ \\
 \hline
 $\pi^{+}$ &  (4.8 $\pm$  1.0)$\times$10$^{-7}$  \\
 \hline
 $\pi^{-}$ &  (4.9 $\pm$  1.0)$\times$10$^{-7}$ \\
 \hline
\end{tabular}
\caption{Particle fluence rate in Ge crystal (thickness 10 cm) at 20 mwe} 
\label{table:fluence rate}
\end{table}

%\end{comment}

\subsection{Systematic Uncertainties and Contributions from Different Processes}
\label{Systematics uncertanity}
The main sources of systematic uncertainty in the activation rates are listed in Table \ref{table:uncertainty_production}. An additional 10\% uncertainty is included to account for statistical uncertainties in the neutron-induced production. The particle flux (including the negative muon stopping rate) and the production cross-sections are treated as independent and combined in quadrature to obtain the systematic uncertainty for each individual process.

The total systematic uncertainty is then obtained by combining the uncertainties from the individual processes, weighted by their relative contributions to the overall production at each depth. The error bars in the production suppression factors shown in Figure \ref{fig:production_suppresion_factor_fit} represent the resulting total uncertainty.

Uncertainties in the sea-level production rates are also included, but are assumed to be uncorrelated with those at shallow depths when evaluating the total uncertainty in the suppression factors, since neutron-induced production contributes only a small fraction of the total production at shallow depths.
\begin{table}[ht!]
\centering
\tiny
\begin{tabular}{|p{1.0cm}|p{2.2cm}|p{1.3cm}|p{2.0cm}|}
\hline
\textbf{Source} & \textbf{Component} & \textbf{Uncertainty (\%)} & \textbf{Comment} \\
\hline
Neutron & $> 10$ MeV neutron flux & $\sim$ 40 & Measurements at SUF \cite{chen1993measurements} \\
\cline{2-4}
 & Cross-sections (Si, Ge, Cu) & 10, 25, 90  & Sea-level measurements \\
\cline{2-4}
 & Composition dependence & $\sim$ 30 & Section \ref{Muon-induced neutron cosmogenic activation} \\
\cline{2-4}
 & Back scattering neutron flux& $\sim$ 30 & Section \ref{back-injected neutrons}\\
 \cline{2-4}
 & Statistical uncertainty in neutron flux & 10 & Section \ref{sec:neutron_flux}\\
\hline
Stopping muon & Muon stopping rate & 35 & Section \ref{Stopping negative muon rate} \\
\cline{2-4}
 & Negative muon capture prob. & $\sim$ 10 & \cite{suzuki1987total} \\
\cline{2-4}
 & Yield per negative muon capture (Si, Ge, Cu) & $\sim$ 25, 50, 200 & Section \ref{Production yield per stopped negative muon} \\
\cline{2-4}
 & mwe depth & $\sim$ 30 & Percent change (For depth $\pm$ 3 mwe) \\
\hline
Photon & Production rate (Si, Ge, Cu) & 100,100,50 & Section \ref{Photo-nuclear activation} \\
\hline
Others & Production rate & 100 & Section \ref{Other processes}\\
\hline
\end{tabular}
\caption{Major systematic uncertainties}
\label{table:uncertainty_production}
\end{table}

\begin{table}[h!]
\centering
\begin{tabular}{|l|l|c|c|c|}
\hline
\multirow{2}{*}{Location} & \multirow{2}{*}{Process} & \multicolumn{2}{c|}{Tritium} & \isotope{Co}{60} \\
\cline{3-5}
 &  & Si & Ge & Cu \\
\hline
\multirow{2}{*}{\shortstack{Sea-level}} 
 & Neutrons & 98$\%$ & 99$\%$ & 93$\%$ \\
 & Others & 2$\%$ & 1$\%$ & 1$\%$  \\
\hline
\multirow{4}{*}{\shortstack{SUF Tunnels A/C \\ 15-20 mwe}} 
 & Stopping neg. muons & 50$\%$ & 38$\%$ & 57$\%$ \\
 & Neutrons & 23$\%$ & 33$\%$  &  15$\%$\\
 & Gammas & 9$\%$ & 1$\%$ &  16$\%$ \\
 & Others & 18$\%$ & 27$\%$ & 12$\%$\\
\hline
\multirow{4}{*}{\shortstack{PNNL SUL \\ $\sim$ 30 mwe}} 
 & Stopping neg. muons & 41$\%$ & 31$\%$  & 50$\%$  \\
 & Neutrons & 28$\%$ & 39$\%$ & 19$\%$  \\
 & Gammas & 10$\%$  & 1$\%$ & 18$\%$ \\
 & Others & 21$\%$ & 29$\%$  & 14$\%$ \\
\hline
\multirow{4}{*}{\shortstack{SLC Adit Storage \\ 50-60 mwe}} 
 & Stopping neg. muons & 37$\%$ & 25$\%$  & 41$\%$ \\
 & Neutrons & 30$\%$ & 42$\%$ & 19$\%$ \\
 & Gammas & 13$\%$ & 2$\%$  & 22$\%$ \\
 & Others & 21$\%$ & 32$\%$  & 18$\%$ \\
\hline
\end{tabular}
\caption{Relative contributions of various physical processes leading to tritium production in Ge and Si and \isotope{Co}{60} production in Cu for the shallow-depth sites considered in this study. The relative contributions at sea level are also shown for comparison.}
\label{tab:relative_contributions}
\end{table}
Based on this study, the uncertainty in \isotope{Co}{60} production in Cu is dominated by the yield from negative muon capture.

\subsection{Production suppression factor vs depth}
\label{Production suppression factor vs depth}
The production suppression factor (defined with respect to the sea-level production rate) as a function of depth is shown in Figure \ref{fig:production_suppresion_factor_fit}. The uncertainties are obtained as discussed in Appendix \ref{Systematics uncertanity}. A simple power-law fit to the data is also shown in the figure. The fit function is $S(h) = a(h/h_0)^{b}$, where $S(h)$ is the suppression factor, $h$ is the depth in mwe, and $h_0$ is the geometric mean of the depths used in the datapoints. Choosing $h_0$ to minimize the correlation leads to a value close to the geometric mean, as expected \cite{montgomery2021introduction} for a power-law fit. The parameters $a$ and $b$ are left free in the fit. Using the geometric mean of the depth reduces the correlation between the fit parameters to less than 10% for all three fits. The $\chi^2/\mathrm{ndf}$ for all fits is $< 1$, but the uncertainties on the parameters are large.

The parameterization is expected to be valid at shallow depths; at greater depths, where activation from stopped negative muons becomes negligible, the curve is expected to fall more steeply.

\begin{figure}[ht!]
% \centering
 
\subfloat[]{
	\label{subfigure:a}
	\includegraphics[width=0.45\textwidth]{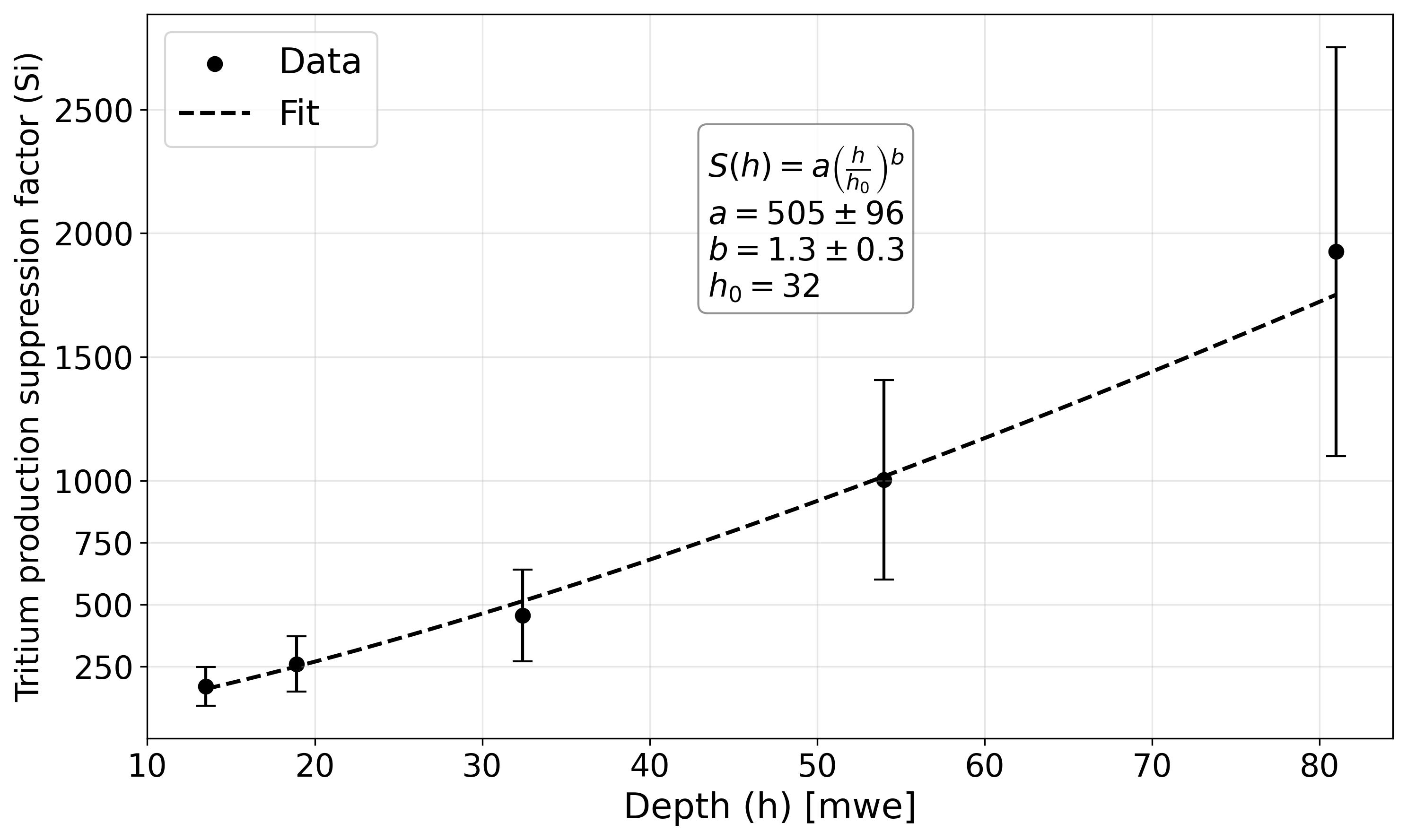}} 
\hspace{0.001\textwidth} % Optional horizontal spacing
\subfloat[]{
	\label{subfigure:b}
	\includegraphics[width=0.45\textwidth]{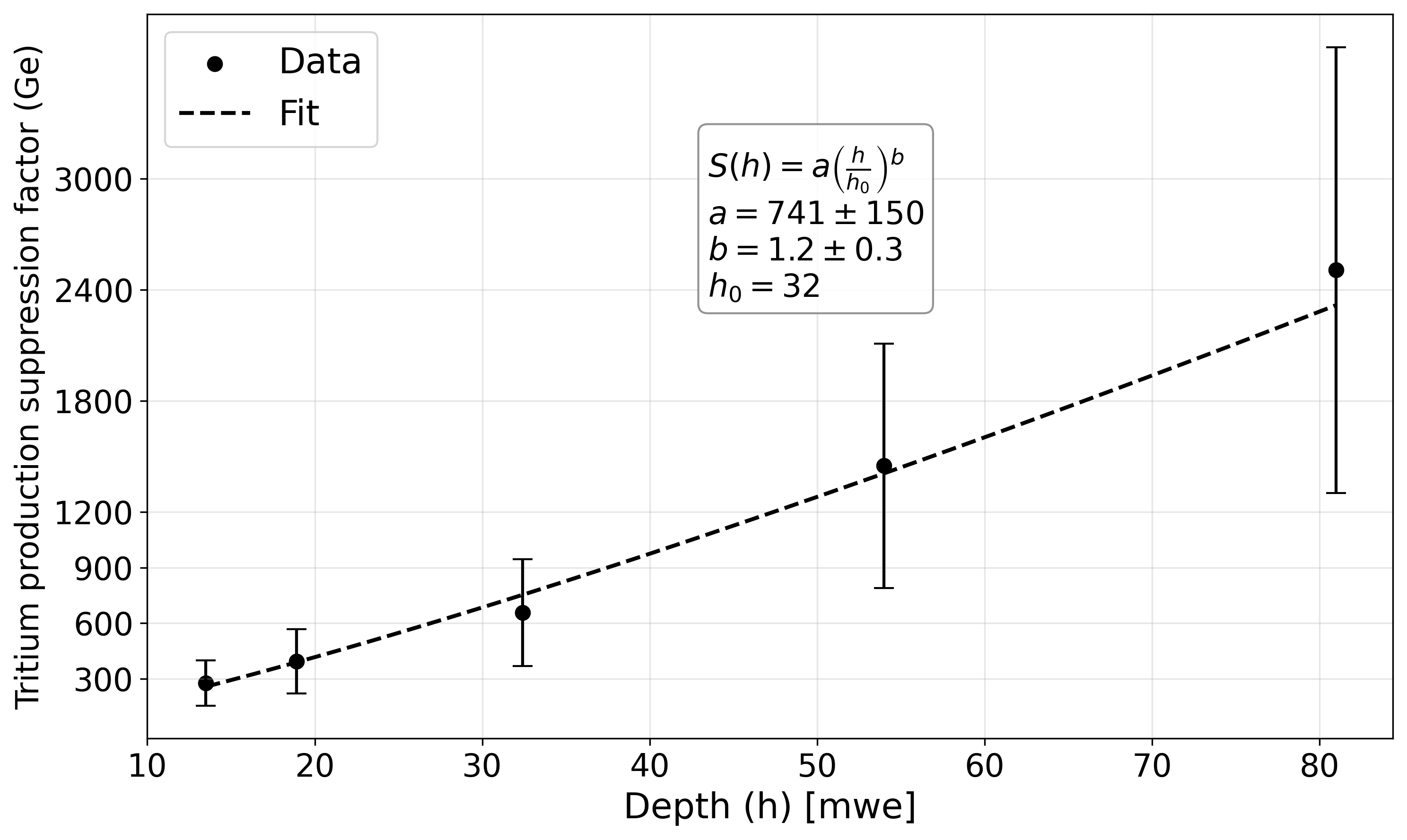}} 
\hspace{0.01\textwidth} % Optional horizontal spacing
\subfloat[]{
	\label{subfigure:c}
	\includegraphics[width=0.45\textwidth]{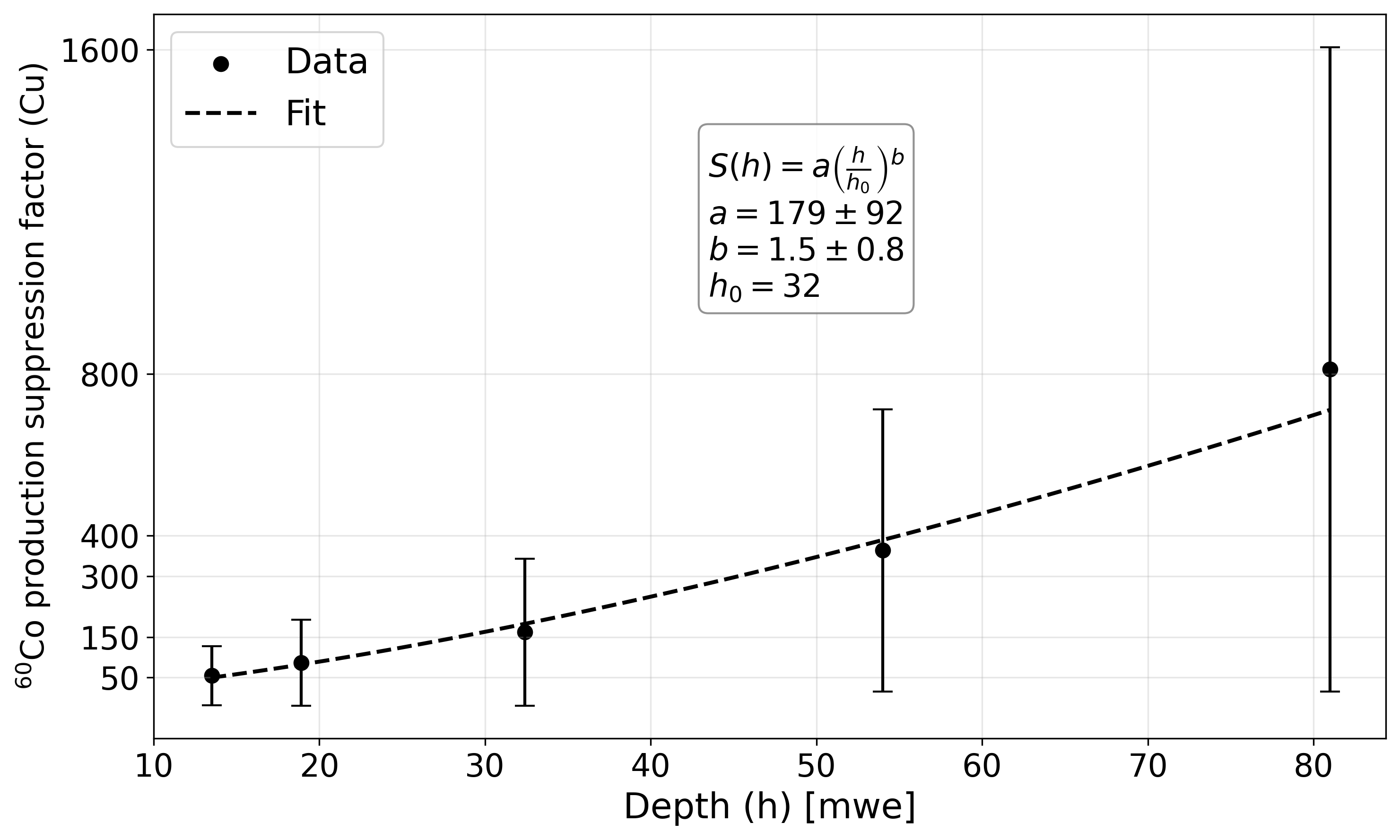}} 
 
\caption{Production suppression factors relative to sea-level production rates for tritium in Si (a) and Ge (b), and \isotope{Co}{60} production in Cu (c), as functions of depth. Power-law fits, with their fit parameters, are also shown in the text boxes.}
 \label{fig:production_suppresion_factor_fit} 
 \end{figure}

\end{document}